\documentclass[a4paper,11pt,superscriptaddress,nofootinbib]{article}
\usepackage{geometry}
\usepackage{amssymb}
\usepackage{amsmath}
\usepackage{graphicx}
\usepackage{dcolumn}
\usepackage{bm}
\usepackage[usenames]{color}
\usepackage[colorlinks=true,linkcolor=blue,citecolor=blue,anchorcolor=green,urlcolor=blue]{hyperref}
\usepackage[percent]{overpic}
\usepackage{tabularx}
\usepackage{multirow}
\usepackage{CJK}
\usepackage{orcidlink}
\usepackage{booktabs}
\usepackage{makecell}
\usepackage{float}
\floatstyle{plaintop}
\restylefloat{table}
\usepackage[symbol]{footmisc}

\renewcommand{\selectlanguage}[1]{}

\def\be{\begin{equation}}
\def\ee{\end{equation}}
\def\bea{\begin{eqnarray}}
\def\eea{\end{eqnarray}}

\def\ba{\begin{aligned}}
\def\ea{\end{aligned}}
\def\mc{\mathcal}

\begin{document}

\begin{center}
\Large{\bf Mesons in a quantum Ising ladder}
\end{center}

\vspace{.2in}
\begin{center}
Yunjing Gao$^1$, Yunfeng Jiang$^{2,3}$\footnote{jinagyf2008@seu.edu.cn}, and Jianda Wu$^{1,4,5}$\footnote{wujd@sjtu.edu.cn}
\\
\vspace{.3in}
\small{
\textit{
1. Tsung-Dao Lee Institute, Shanghai Jiao Tong University, Shanghai, 201210, China\\
2. School of Phyiscs and Shing-Tung Yau Center,\\ Southeast University, Nanjing 210096, China\\
3. Peng Huanwu Center for Fundamental Theory, Hefei, Anhui 230026, China\\
4. School of Physics \& Astronomy,\\ Shanghai Jiao Tong University, Shanghai, 200240, China\\
5. Shanghai Branch, Hefei National Laboratory, Shanghai 201315, China\\
}}

\end{center}

\begin{abstract}
    When two transverse-field Ising chains (TFICs) with magnetic order are coupled, the original free excitations become confined, giving
rise to meson-like bound states. In this work, we study such bound states systematically.
    The mesons are characterized by their fermion number parity and
chain-exchanging properties, which lead to distinct sets of mesonic states. 
    The meson masses are determined by solving the Bethe-Salpter equation. An interesting observation is
    the additional degeneracy in the chain-exchanging odd sectors. 
    Beyond the two particle approximation, we exploit the truncated free fermionic space approach to calculate the spectrum numerically. 
    Corrections to the meson masses are obtained, 
    and the degeneracy is further confirmed. 
    The characterization and degeneracy can be 
    connected to the situation when
    each chain is tuned to be quantum critical, 
    where the system is described by the Ising$_h^2$ integrable model,
    a sine-Gordon theory with $\mathbb{Z}_2$ orbifold.   
    Here we establish a clear correspondence between the particles in the bosonized form and their fermionic counterparts.
    Near this point,
    the stability of these particles is analyzed using the form factor perturbation scheme, 
    where four particles are always present. Additionally, we calculate the evolution of the dominant
    dynamical structure factor for local spin operators,
    providing further insight into the low-energy excitations and their role in the system's behavior.
    The two-particle confinement framework as well as the parity classifications may inspire the study for other coupled bi-partite systems.
\end{abstract}

\section{Introduction}
Contrary to the electromagnetic interaction that decays with distance,
quark confinement in quantum chromodynamics (QCD) provides another scenario for bound states:
the attractive force due to the strong interaction 
tends to increase (within a certain scale) as two particles are pulled apart \cite{PhysRevD.10.2445}.
It means that the constituent particles (quarks and gluons) cannot be solely observed but 
always appear as composite particles, 
\textit{e.g.}, baryons, mesons and glueballs. 
Understanding color confinement in QCD and giving a rigorous proof to the mass gap of the quantum Yang-Mills theory are among the most profound open questions in quantum field theory. In 1+1 dimensions, however, the problem is simplified significantly and the meson spectrum for 2d QCD has been obtained already in 1978 by 't Hooft \cite{tHooft:1974pnl} (see \cite{Abdalla:1995dm,Fateev:2009jf,Ambrosino:2023dik,Ambrosino:2024prz,Litvinov:2024riz} for more recent developments).\par

Confinement is not unique to QCD, but is a more universal phenomena in physics. 
In fact, without delving into the strong interaction region,
similar mechanism has also been revealed in low-dimensional statistical mechanics and condensed matter systems. 
In the pioneering work of McCoy and Wu \cite{PhysRevD.18.1259}, it was already pointed out that as one varies the two parameters of the model (temperature and magnetic field), 
the spectrum of stable particles of the corresponding field theory can contain an infinite tower of mesons formed by confined `quarks'. 
Such a physical scenario was confirmed and extended in a series of beautiful works termed `Ising spectropy' \cite{Fonseca2003,zam2006,Zamolodchikov:2013ama,Xu:2022mmw,Xu:2023nke}. 
In integrable quantum field theories (IQFT), 
soliton confinement is intimately related to integrability breaking. 
Turning on an integrability-breaking perturbation for an IQFT typically induces non-local interactions among the stable particles that leads to confinement \cite{Mussardo:2010sy}. This mechanism has been confirmed in a number of models including $Q$-state Potts model \cite{DELFINO2008265}, tricritical Ising model \cite{Lencses:2021igo} and double period sine-Gordon model \cite{DELFINO1998,Roy:2023byz,Rutkevich:2023lzj}.
Since most of the aforementioned QFTs typically describe the continuum limit of underlying spin models \cite{PhysRevD.18.1259,PhysRevD.23.1862,PhysRevLett.77.2790,DELFINO2008265,PhysRevLett.83.3069,Ramos_PRB_2020},
not only they could be studied theoretically, 
but concrete connections between the composite excitations and several quantum magnets have been discussed \cite{Rutkevich_2010,jianda_E8_2014}. 
Experimental efforts have also been made to detect the excitations through inelastic neutron scattering experiments and terahertz measurement \cite{coldea_quantum_2010,E8,zhang_e8_observation_2020,wang_tfic_quantum_2018,PhysRevB.96.054423,WANG20242974}. 
Non-equilibirum dynamics in these systems is further investigated \cite{Kormos2017,PhysRevB.110.195101,geometry}.
In recent years, it has also been further discovered that confinement can drastically change the non-equilibrium dynamics in quantum spin systems, opening up exciting new avenues to explore both in theory \cite{Kormos2017,Pomponio:2024aiy} and in experiments \cite{SciRep2021,Tan:2019kya}.

A prototype of low-dimensional quantum many-body systems which exhibits confinement is the two-dimensional (2D) Ising model 
with the presence of magnetic field along Ising spin direction \cite{PhysRevD.18.1259}.
The Ising field theory emerges as the scaling limit of the 2D lattice Ising model,
with the critical point ($T=T_c$ and zero field) associated with 
the central charge $c=1/2$ conformal field theory (CFT) \cite{ZamE8}.
The theory contains two relevant operators, 
spin density $\sigma$ and energy density $\epsilon$,
with conformal dimensions $(\frac{1}{16},\frac{1}{16})$ and $(\frac{1}{2},\frac{1}{2})$, respectively \cite{zamolodchikov1987higher,Zamolodchikov:1989hfa}.
Relevant perturbation at this point leads to the Hamiltonian
\be
H_{\text{Ising}}(\tau,h)=H_{c=1/2}+\tau\int dx \epsilon(x)+h\int dx \sigma(x),
\label{eq:tfic}
\ee
with $H_{c=1/2}$ denoting the $c=1/2$ CFT. Here $\tau\simeq T-T_c$ and $h$ is the rescaled magnetic field.
Field theory in Eq.~\eqref{eq:tfic} effectively describes the continuum limit of the transverse field Ising chain (TFIC) perturbed by a longitudinal field,
with the lattice Hamiltonian 
\be
H_{\text{TFIC}_h}(g,H)=J\sum_{j=1}^{N-1}\sigma_j^{z}\sigma_{j+1}^{z}+J\sum_{j=1}^N\left(g\sigma_j^x+H\sigma_j^z\right).
\label{eq:QIC}
\ee
Here $\sigma_{j}^{\beta}$ ($\beta =z,x$) denotes $\beta$-component of the Pauli matrix at site $j$, 
$J$ is the Ising coupling strength between nearest Ising spins, 
which serves as an overall energy scale;
$g$ and $H$ are the reduced coupling strength of transverse and longitudinal field, respectively.
When $H=0$,
Eq.~\eqref{eq:QIC} encounters a quantum phase transition at $g=1$.
$T<T_c$ corresponds to $g<1$, 
which is referred to as the Ising ordered phase \cite{sachdev_2011}.
$H_{\text{TFIC}_h}(g,0)$ can be exactly mapped to a free fermion model \cite{pfeuty_one-dimensional_1970},
with the basic excitation effectively understood as domain wall when $g<1$. 
The theory is integrable for $H=0$. For $H\ne 0$ and $T\ne T_c$, integrability is broken and the spectrum contains an infinite tower of mesons.
An analytical approach to understand the meson spectrum has been developed in \cite{zam2006},
which mainly consists of two particles confined by an effectively long-range force provided by the magnetic field.

Due to its central role in quantum phase transition, understanding meson spectrum for Ising field theory is physically interesting and important. 
In this work, we study an equally important prototypical model which describes two copies of $H_{\text{Ising}}(\tau,0)$ that are weakly coupled to each other via the two spin fields. 
This model describes the continuum limit of the quantum Ising ladder on the lattice.
Integrablilty of the coupled system with $\tau=0$ 
emerges as the Ising$_h^2$ integrable field theory (IIFT) \cite{coupleCFT},
which can be organized as a sine-Gordon type model with $\mathbb{Z}_2$ orbifold \cite{QFTising,2isingboson}.
The IIFT contains six breathes ($B_1, \;\cdots,\; B_6$) and two solitons ($A_+,\; A_-$).
It has been discovered that $B_1$, $B_3$, $B_5$ cannot be directly excited from the ground state \textit{through (quasi-)local spin operations}, 
nor can they spontaneously decay to the ground state through vacuum fluctuations when
they are prepared \cite{dark}.
This calls for a better physical understanding of the particle classifications provided by the sine-Gordon framework.
It further raises the question of whether the particles and aforementioned properties hold for $\tau\neq0$.
At the same time,
quantum magnets that can effectively realize the Ising ladder as well as the detection of $B_1$ partilce have been proposed \cite{Xning,thermal}.
Furthermore,
as the structure and interaction are relatively simple in the lattice model, it may be possible to directly simulate it in finite-sized systems, such as Rydberg arrays \cite{rydberg}. 
As we will show in this paper, for $\tau\ne 0$, 
the model is no longer integrable and exhibits confinement. 
Compared to the Ising chain, the meson structure is richer. 
The fundamental excitation on a single chain is a single domain wall. 
Now, 
with two chains in the ladder system, 
confinement of fundamental excitations leads to two classes of mesons. 
One is formed by confining two domain walls on the same chain, we call this the \textit{intrachain} mesons. 
The other type of meson is called \textit{interchain} meson.
Confinement for the latter type is new compared to the Ising chain and is led by the
mismatch in the domain wall positions between the two chains. 
One of the main results of this paper is a quantitative determination of the spectrum of these two types of mesons.

The structure of the paper is as follows. 
In Sec.~\ref{sec:model}
a linear-potential approximation is presented for simulating the confinement,
which gives rise to two sets of mass solutions for the mesons. 
This calculation, albeit simple, offers the basic physical picture and qualitative results.
To obtain a more accurate meson spectrum, we establish
a Bethe-Salpeter (BS) type equation \cite{zam2006} in Sec.~\ref{sec:BS},
we find four sets of mass solutions, 
characterized by fermion number parity ($\mc{N}$) on each chain,
and sign change of the wavefunction 
after exchanging the two chains ($+1$ or $-1$, referred to as $U_{\text{ex}}$ even and odd).
Analytical
solution of the BS equation is obtained in the weak coupling region by perturbation theory,
and the mass spectrum in the full parameter space is numerically solved.
An interesting observation from the analysis of BS equation is the degeneracy between $\mc{N}$-odd and $\mc{N}$-even sectors with $U_{\text{ex}}$ odd,
where the BS equations take the same form.
To further confirm the meson confinement picture, 
in Sec.~\ref{sec:tffsa}, we apply the truncated free fermionic space approach (TFFSA) to determine the spectrum numerically. 
The results from the TFFSA is consistent with the ones from BS equation, 
in particular the degeneracy is further confirmed beyond the two-particle subspace projection. 
In addition,  the BS equation results are improved by a renormalized tension factor obtained from TFFSA.
Connection between the aforementioned classification and the Ising$_h^2$ integrable model is 
discussed through the truncated CFT calculation in Sec.~\ref{sec:integrable},
where the charge conjugation parity of the breathers is identified as fermion number parity,
and the solitons and breathers are found in $U_{\text{ex}}$-odd and even sectors, respectively.
Particle stability near the integrable point is discussed using the form factor perturbation scheme \cite{DELFINO1996,DELFINO2006},
where the evolution of the dynamical spectrum for single particles is also presented.

\section{The Ising ladder}
\label{sec:model}
\subsection{The model}
Consider the field theory which describes two coupled TFICs in the scaling limit,
the Hamiltonian reads
\be \ba
&H_{\text{II}}=H_{\text{Ising}}^{(1)}(\tau,0)+H_{\text{Ising}}^{(2)}(\tau,0)+\lambda\int dx\sigma^{(1)}(x)\sigma^{(2)}(x),
\label{eq:action}
\ea\ee
where $H_{\text{Ising}}$ is introduced in Eq.~\eqref{eq:tfic},
$(1,2)$ are the chain labels and 
$\lambda$ refers to the interchain coupling strength.
Eq.~\eqref{eq:action} is the continuum limit of the transverse-field Ising ladder
\be
H_{\text{ladder}}=J\sum_{j=1}^{N-1}\left(\sigma_j^{z(1)}\sigma_{j+1}^{z(1)}+\sigma_j^{z(2)}\sigma_{j+1}^{z(2)}\right)+J\sum_{j=1}^Ng\left(\sigma_j^{x(1)}+\sigma_j^{x(2)}\right)+J_i\sigma_j^{z(1)}\sigma_j^{z(2)},
\label{eq:lattice}
\ee
$J$ and $g$ are the same as in Eq.~\eqref{eq:QIC},
and $J_i$ is the reduced interchain coupling strength.
The IIFT emerges when $H^{(1,2)}_{\text{Ising}}(\tau,0)$ are tuned to $\tau=0$ ($g=1$).
Hamiltonian \eqref{eq:action} can not be solved exactly in general.

Starting from the discrete model $H^{(a)}_{\text{TFIC}_h}(g,0)$,
spins on each chain can be constructed from a set of fermion creation and annihilation operators $d_j^{(a)\dagger}$, $d_j^{(a)}$ \cite{pfeuty_one-dimensional_1970,2isingboson,Lieb1961} as
\be
\sigma_j^{x(a)}=2d_j^{(a)\dagger}d_j^{(a)}-1,\quad 
\sigma_j^{z(a)}=(d_j^{(a)\dagger}+d_j^{(a)})\mu^{(a)}_j,\quad
\mu_j^{(a)}=\prod_{l<j}\sigma_l^{x(a)}=e^{i\pi\sum_{l<j}d_l^{(a)\dagger}d^{(a)}_l}.
\label{eq:op}
\ee
It is convenient to introduce the fermions in the momentum space as
$d_k^{(a)}=\sum_{j=1}^Nd_j^{(a)}e^{-ik(j\tilde{a})}/\sqrt{N}$,
with $k$ being the lattice momentum.
$H_{\text{TFIC}_h}^{(a)}(g,0)$ can be diagonalized 
by performing the Bogoliubov transformation,
\be
\gamma_k^{(a)}=u_kd_k^{(a)}-iv_kd_{-k}^{(a)\dagger},\quad 
\gamma_k^{(a)\dagger}=u_kd_k^{(a)\dagger}+iv_kd_{-k}^{(a)},
\ee
where $u_k=\cos(\theta_k/2)$,
$v_k = \sin(\theta_k/2)$,
with $\cos\theta_k=2J(g-\cos k)/\epsilon_k$ and single particle dispersion $\epsilon_k=2J\sqrt{1-2g\cos k+g^2}$.
$H_{\text{TFIC}_h}^{(a)}(g,0)$ is further organized as 
\be
H_{\text{TFIC}_h}^{(a)}(g,0)=\sum_{k}\epsilon_k\left(\gamma_k^{(a)\dagger}\gamma_k^{(a)}-\frac{1}{2}\right).
\label{eq:lat}
\ee

Real spinors $\psi^{(a)\intercal}(j)=(\psi^{(a)}_L(j),\psi^{(a)}_R(j))$ are introduced through 
\be
\psi^{(a)}_L(j)=\frac{(-1)^j}{\sqrt{2\tilde{a}}}(d_j^{(a)\dagger}e^{-i\pi/4}+d_j^{(a)}e^{i\pi/4}),
\quad 
\psi^{(a)}_R(j)=\frac{(-1)^j}{\sqrt{2\tilde{a}}}(d_j^{(a)\dagger}e^{i\pi/4}+d_j^{(a)}e^{-i\pi/4}).
\label{eq:phi}
\ee
Taking the lattice spacing $\tilde{a}\to0$ and denoting $x=j\tilde{a}$,
such that $\psi^{(a)\intercal}(x)=(\psi^{(a)}_L(x),\psi^{(a)}_R(x))$.
$H_{\text{TFIC}_h}(g,0)$ in the continuum limit
is organized as the free fermionic model 
\be 
H_{\text{Ising}}^{(a)}(\tau,0)=\int dx\left[\psi(x)\left(-i\tilde{a}\gamma^5\frac{\partial}{\partial x}\psi(x)\right)+(g-1)\psi(x)\gamma^0\psi(x)\right],
\label{eq:free}
\ee
with $\gamma^0=\sigma^y$ and $\gamma^5=\sigma^z$.
Parallel to Eq.~\eqref{eq:lat},
in second quantization, $H_{\text{Ising}}^{(a)}(\tau,0)$
takes the form of
\be 
H^{(a)}_{\text{Ising}}(\tau,0)=E_{\text{vac}}+\int_{-\infty}^{\infty}\frac{dp}{2\pi}\omega(p)c^{(a)\dagger}(p)c^{(a)}(p),\,
\label{eq:spinor}
\ee
where $E_{\text{vac}}$ is the vacuum energy. 
The fermionic operators $c^\dagger$, $c$ satisfy $\{c^{(a)}(p),c^{(a)\dagger}(p^{\prime})\}=\delta(p-p^{\prime})$, 
$\{c^{(a)}(p),c^{(a)}(p^{\prime})\}=\{c^{(a)\dagger}(p),c^{(a)\dagger}(p^{\prime})\}=0$,
and anticommute between different chains, with momenta $p$ and $p^{\prime}$.
The energy for a single fermion is given by $\omega(p) = \sqrt{m^2+p^2}$ with fermion mass $m\equiv2\pi\tau$.

\begin{figure}[htp]
    \centering
    \includegraphics[width=0.7\textwidth]{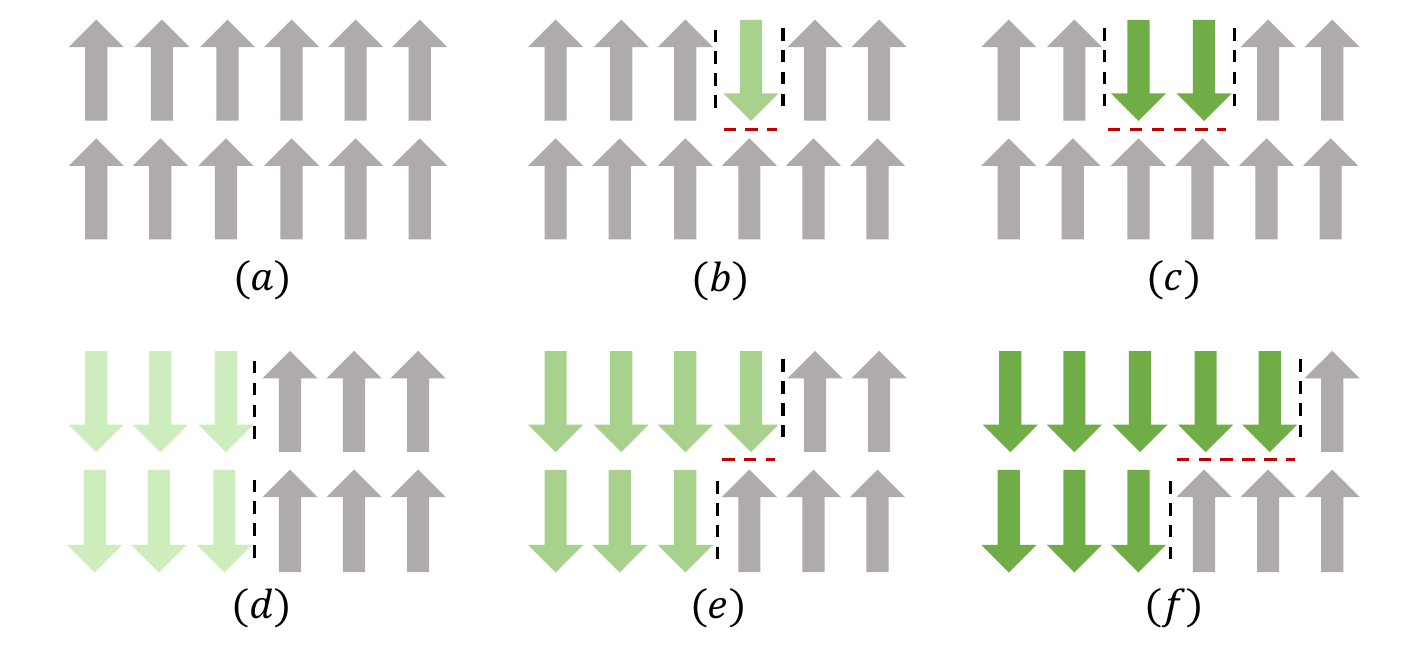}
    \caption{Illustration on formation of intrachain meson ($a$-$c$) and interchain meson ($d$-$f$) ($J<0$). 
    Black dashed lines represent domain walls in each chain (intrachain domain walls) and red ones denote domain walls in between (intercahin domain walls).
    Increasing the horizontal distance between the intrachain domain walls leads to higher energy, illustrated with darker green.}
    \label{fig:conf}
\end{figure}

In the Ising ordered phase of $H_{\text{TFIC}_h}^{(a)}(g,0)$,
elementary excitations are identified as single fermions, which
can be pictorically understood as a domain wall.
In the presence of interchain coupling, two anti-parallel spins at the same site between the two chains will raise the energy.
Increasing the distance between the two anti-parallel spins is energetically unfavorable, leading to the confinement of pair of domain walls as shown in Fig.~\ref{fig:conf}.

\subsection{Linear potential approximation}
The confinement of two fermions can be qualitatively described
by a linear potential $V(x)=f_0|x|$,
as the energy increases with the distance $|x|$ between the two domain walls.
Here $f_0$ plays the role of the string tension.
In the center of mass frame of the meson, 
we have 
\be 
-\frac{1}{2\mu}\frac{d^2\Psi}{dx^2}+f_0|x|\Psi=E\Psi,
\quad 
\Psi(|x|\to \infty)=0,
\label{eq:sch}
\ee 
where $\Psi(x)$ is the meson wavefunction and $\mu=m/2$ is the reduced mass.
Eq.~\eqref{eq:sch} takes the same form as that for the Ising chain in Eq.~\eqref{eq:tfic} \cite{PhysRevD.23.1862, zam2006},
while the initial conditions differ due to the source of the constituent particles.
Eq.~\eqref{eq:sch} can be solved by the Airy function
\be 
\Psi = \text{Ai}(\mu^{1/3}f_0^{1/3}2^{1/3}(x-f_0^{-1}E)),\quad x\geq0.
\ee 
For the intrachain meson, \emph{i.e.},
the two particles confined in the same chain,
we require
\be
\Psi(x=0)=0
\label{eq:bc1}
\ee
due to Pauli exclusion principle.
This leads to 
\be
E_n=z_n(2\mu)^{-1/3}f_0^{2/3},\quad n=1,2,\cdots
\ee 
with $z_n$ obtained from $\text{Ai}(-z_n)=0$.
Combining $E_n$ with the rest mass,
total energies in this frame, or equivalently
the meson masses are given by 
\be
M_n=2m+z_nm^{-1/3}f_0^{2/3}.
\label{eq:sol1}
\ee

For the interchain meson composed by two particles from different chains,
they can occupy the same position along the chain direction.
Furthermore,
since $H_{\text{II}}$ (Eq.~\ref{eq:action}) is invariant under the exchange of chain $1$ and $2$ ($U_{\text{ex}}$),
the Hilbert space can be divided into two sectors:
one that remains invariant (referred to as even) and one that gains a minus sign (odd) under the action of $U_{\text{ex}}$.
By assigning positions $x_1$ and $x_2$ to the two fermions from different chains,
we can construct meson wavefunctions as $\widetilde{\Psi}^{\pm}(x_1;x_2)=[\widetilde{\Psi}(x_1;x_2)\pm \widetilde{\Psi}(x_2;x_1)]/2\equiv \widetilde{\Psi}^{\pm}(x_1-x_2)$, where $\widetilde{\Psi}^+$ and $\widetilde{\Psi}^-$ satisfy Eq.~\eqref{eq:sch}, belonging to $U_{\text{ex}}$-even and -odd sectors respectively.
Since $\widetilde{\Psi}^-(x_1;x_1)=0$,
it follows condition Eq.~\eqref{eq:bc1}, 
leading to the same solutions as in Eq.~\eqref{eq:sol1}.
$\widetilde{\Psi}^{+}(x_1;x_2)=\widetilde{\Psi}^{+}(x_2;x_1)$
leads to the result
$d\widetilde{\Psi}^+(x)/dx=-d\widetilde{\Psi}^{+}(-x)/dx$, namely
\be 
\left.\frac{d\widetilde{\Psi}^+(x)}{dx}\right|_{x=0}=0.
\ee
This indicates that when the domain walls from the two chains 
meet at the same position, the energy is minimized when they do 
not have relative motion.
Correspondingly, 
by introducing $\tilde{z}_n$ through Ai$^{\prime}(-\tilde{z}_n)=0$,
we have the other set of meson masses
\be
M_n=2m+\tilde{z}_nm^{-1/3}f_0^{2/3},\quad n=1,2,\cdots
\label{eq:lp}
\ee 
Already from this simple analysis, 
we have gained two important features of the meson spectrum in the Ising ladder. 
Firstly, in addition to the Ising-like intrachain mesons, 
there are additional types of mesons which originates from the confinement of interchain excitations. 
Additionally, interchain mesons can be categorized into $U_{\text{ex}}$-even or -odd sector, where
$U_{\text{ex}}$-odd sector exhibits a degeneracy with the spectrum of intrachain mesons.

\section{The Bethe-Salpeter equation}
\label{sec:BS}
In order to obtain more accurate results for the meson spectrum beyond the linear potential picture, we adopt a more systematic approach based on two-particle approximation. 
The relativistic two-particle bound state equation is known as the Bethe-Salpeter equation,
which has been applied to the analysis of Ising field theory \cite{zam2006}. 
In this section, we derive the BS-like equations for the Ising ladder system and find the meson spectrum by solving these equations.

\subsection{The two-particle subspace}
Following the domain wall picture,
if the fermion number parities of the two chains differ,
spins over a large range are antiparallel between the two chains,
which are not among the low-energy configurations we focus on.
We can formally construct the eigenstates for the Hamiltonian in Eq.~\eqref{eq:action} as $|\Phi\rangle=|\Phi^{(2)}\rangle+|\Phi^{(4)}\rangle+|\Phi^{(6)}\rangle+\cdots$ similar to \cite{zam2006},
with upper index denoting the total fermion numbers.
The two particle states can be constructed as
\be \ba
|\Phi^{(2)}\rangle=&\frac{1}{2}\int_{-\infty}^{\infty}\frac{dq_1}{2\pi}\frac{dq_2}{2\pi}\Psi^{(2,0)}(q_1,q_2)|q_1,q_2\rangle_1\otimes|0\rangle_2\\
+&\frac{1}{2}\int_{-\infty}^{\infty}\frac{dq_1}{2\pi}\frac{dq_2}{2\pi}\Psi^{(0,2)}(q_1,q_2)|0\rangle_2\otimes|q_1,q_2\rangle_1\\
+&\int_{-\infty}^{\infty}\frac{dq_1}{2\pi}\frac{dq_2}{2\pi}\Psi^{(1,1)}(q_1;q_2)|q_1\rangle_1\otimes|q_2\rangle_2,
\label{eq:wf}
\ea \ee
with $q_1$ and $q_2$ denoting the fermion momenta.
Here $|p,q\rangle_{a}=c_p^{(a)\dagger}c_q^{(a)\dagger}|0\rangle_{a}$, 
$_a\langle0|c^{(a)}_qc^{(a)}_p={_a\langle} q,p|$, 
and $|0\rangle_{a}$ denotes the vacuum for chain $a=1,2$. 
$\Psi^{(n_1,n_2)}$ are the coefficients to be determined,
with $n_1$ and $n_2$ denoting the fermion number for chain $1$ and $2$.
Eq.~\eqref{eq:wf} includes both cases: where
the two domain walls are either within a single chain or spread into two chains,
leading to intrachain and interchain mesons, respectively.

Acting the Hamiltonian Eq.~\eqref{eq:action} on the above two-particle state,
we have 
\be 
H_{\text{II}}|\Phi\rangle = (2E_{\text{vac}}+\Delta E)|\Phi\rangle,
\label{eq:Hpsi}
\ee
with $\Delta E$ to be determined.
Projecting Eq.~\eqref{eq:Hpsi} on the $\mc{N}$-even state $\langle p_1,p_2|_1\otimes\langle0|_2$ leads to 
\be\ba
&-(\omega(p_1)+\omega(p_2)-\Delta E)\Psi^{(2,0)}(p_1,p_2)\\
=&-\frac{\lambda}{2}\int_{-\infty}^{\infty}\frac{dq_1}{2\pi}\frac{dq_2}{2\pi}\left(\Psi^{(2,0)}(q_1,q_2)_1\langle p_1,p_2|\sigma^{(1)}(0)|q_1,q_2\rangle_1\overline{\sigma}\right.\\
&+\left.\Psi^{(0,2)}(q_1,q_2)_1\langle p_1,p_2|\sigma^{(1)}(0)|0\rangle_1\,_2\langle0|\sigma^{(2)}|q_1,q_2\rangle_2\right)2\pi\delta(p_1+p_2-q_1-q_2),
\label{eq:bsori1}
\ea\ee
as $\langle q_1\cdots q_n|\sigma|p_1 \cdots p_m\rangle=0$ for $m+n$ being an odd number.
Vacuum expectation value is denoted as $\langle\sigma\rangle=\overline{\sigma}$. 
Projection on the other two types of states can be done similarly, leading to
\be\ba
&-(\omega(p_1)+\omega(p_2)-\Delta E)\Psi^{(0,2)}(p_1,p_2)\\
=&-\frac{\lambda}{2}\int_{-\infty}^{\infty}\frac{dq_1}{2\pi}\frac{dq_2}{2\pi}\left(\Psi^{(0,2)}(q_1,q_2)_2\langle q_1,q_2|\sigma^{(2)}(0)|p_1,p_2\rangle_2\overline{\sigma}\right.\\
+&\left.\Psi^{(2,0)}(q_1,q_2)_1\langle q_1,q_2|\sigma^{(1)}(0)|0\rangle_1\,_2\langle0|\sigma^{(2)}|p_1,p_2\rangle_2\right)2\pi\delta(p_1+p_2-q_1-q_2),\\
\,\\
&(\omega(p_1)+\omega(p_2)-\Delta E)\Psi^{(1,1)}(p_1;p_2)\\
=&-\frac{\lambda}{2}\int_{-\infty}^\infty\frac{dq_1}{2\pi}\frac{dq_2}{2\pi}\Psi^{(1,1)}(q_1;q_2) _1\langle q_1|\sigma^{(1)}|p_1\rangle_1\,_2\langle q_2|\sigma^{(2)}|p_2\rangle_2.
\label{eq:bsori2}
\ea\ee

Here we introduce the notations
\be\ba
\langle p_1,p_2|\sigma(0)|q_1,q_2\rangle\langle0|\sigma(0)|0\rangle&=4\overline{\sigma}^2\mc{G}_1(p_1,p_2|q_1,q_2),\\
\langle p_1,p_2|\sigma(0)|0\rangle \langle0|\sigma(0)|q_1,q_2\rangle&=4\overline{\sigma}^2\mc{G}_2(p_1,p_2|q_1,q_2),\\
\langle p_1|\sigma(0)|q_1\rangle\langle q_1|\sigma(0)|q_2\rangle&=2\overline{\sigma}^2\mc{G}_3(p_1,p_2|q_1,q_2),
\ea\ee
and the following convolution
\be 
\mc{G}\star\Psi(p_1,p_2)\equiv\int\frac{dq_1}{2\pi}\frac{dq_2}{2\pi}2\pi\delta(p_1+p_2-q_1-q_2)\mc{G}(p_1,p_2|q_1,q_2)\Psi(q_1,q_2).
\ee
By constructing $
\Psi_{\pm}^{(2)}(p_1,p_2)=\frac{1}{2}\left(\Psi^{(2,0)}(p_1,p_2)\pm\Psi^{(0,2)}(p_1,p_2)\right)
$,
three decoupled BS equations can be obtained from Eqs.~\eqref{eq:bsori1} and \eqref{eq:bsori2} as
\be\ba
-(\omega(p_1)+\omega(p_2)-\Delta E)\Psi^{(2)}_+(p_1,p_2)&=-2\overline{\sigma}^2\lambda(\mc{G}_1+\mc{G}_2)\star\Psi_+^{(2)}(p_1,p_2),\\
-(\omega(p_1)+\omega(p_2)-\Delta E)\Psi^{(2)}_-(p_1,p_2)&=-2\overline{\sigma}^2\lambda(\mc{G}_1-\mc{G}_2)\star\Psi_-^{(2)}(p_1,p_2),\\
(\omega(p_1)+\omega(p_2)-\Delta E)\Psi^{(1,1)}(p_1;p_2)&=-2\overline{\sigma}^2\lambda\mc{G}_3\star\Psi^{(1,1)}(p_1;p_2).
\label{eq:BS1}
\ea\ee
The kernels used in Eq.\eqref{eq:BS1} have been derived in \cite{Berg_1979,BABELON1992113} and summarized in App.~\ref{app:BSkernel},
which are given by
\be\ba
&\mathcal{G}_1(p_1,p_2|q_1,q_2)=
\frac{-1}{4\sqrt{\omega(p_1)\omega(p_2)\omega(q_1)\omega(q_2)}}\left[\frac{\omega(p_1)+\omega(q_2)}{p_1-q_2}\frac{\omega(p_2)+\omega(q_1)}{p_2-q_1}\right.\\&\qquad\qquad\qquad\qquad\left.-\frac{\omega(p_1)+\omega(q_1)}{p_1-q_1}\frac{\omega(p_2)+\omega(q_2)}{p_2-q_2}+\frac{p_1-p_2}{\omega(p_1)+\omega(p_2)}\frac{q_1-q_2}{\omega(q_1)+\omega(q_2)}\right],\\
&\mathcal{G}_2(p_1,p_2|q_1,q_2)=
\frac{-1}{4\sqrt{\omega(p_1)\omega(p_2)\omega(q_1)\omega(q_2)}}\frac{p_1-p_2}{\omega(p_1)+\omega(p_2)}\frac{q_1-q_2}{\omega(q_1)+\omega(q_2)},\\
&\mathcal{G}_3(p_1;p_2|q_1;q_2)=
\frac{-1}{2\sqrt{\omega(p_1)\omega(p_2)\omega(q_1)\omega(q_2)}}\frac{\omega(p_1)+\omega(q_1)}{p_1-q_1}\frac{\omega(p_2)+\omega(q_2)}{p_2-q_2}.
\ea\ee

\subsection{Infinite momentum limit}
In an interacting theory, 
focusing on the two-particle sector while neglecting higher particle contributions generally breaks Lorentz invariance. 
To restore relativistic invariance, specific multi-particle corrections have to be taken into account. 
However, these corrections disappear in the limit $P\to\infty$ where $P$ is the total momentum. 
This argument was first put forward for the 2d QCD \cite{Bars:1977ud} and later generalized to the IFT \cite{zam2006}. 
We adopt the same strategy here and derive the BS equations by taking the infinite momentum limit of \eqref{eq:BS1}.\par
We first reorganize Eq.~\eqref{eq:BS1} in the center of mass frame.
The meson wave functions with total momentum $P$ can be expressed as 
\be\ba 
\Psi_{\pm}^{(2)}(p_1,p_2)&=2\pi \delta(p_1+p_2-P)\Psi_{P,\pm}^{(2)}(p_1-\tfrac{P}{2}),\\
\Psi^{(1,1)}(p_1;p_2)&=2\pi \delta(p_1+p_2-P)\Psi_{P}^{(1)}(p_1-\tfrac{P}{2}),\\
\ea\ee 
with $\Psi_{P,\pm}^{(2)}(p)=-\Psi_{P,\pm}^{(2)}(-p)$.
Moreover,
all the $\Psi_P(p)$ are assumed to be normalizable that when $P\to\infty$,
$\Psi_P(p)\sim O(1)$ for $|2p/P|<1$ and $\Psi_P(p)\sim O(P^{-2})$ for $|2p/P|>1$ \cite{zam2006}.
Energy of meson with mass $M$ should take the relativistic form $\Delta E=\sqrt{M^2+P^2}$.

Working in the $P\to\infty$ limit,
we have the following expansions when $|u|<1$ ($u\equiv 2p/P$)
\be\ba 
\omega\left(\frac{P}{2}(1+u)\right)&\simeq \frac{|P|}{2}(1+u)+\frac{m^2}{|P|(1+u)},\\
\Delta E&\simeq |P|+\frac{M^2}{2|P|}.
\ea\ee 
By keeping to $O(P)$,
Eq.~\eqref{eq:BS1} can be expressed in terms of the reduced momenta $v=2q/P$ and $u$ as
\be
\ba 
-\left(\frac{m^2}{1-u^2}-\frac{M^2}{4}\right)\Psi^{(2)}_{+}(u)&=f_0\int_{-1}^1\frac{dv}{2\pi}\frac{1}{\sqrt{(1-u^2)(1-v^2)}}\left[\frac{1+uv}{(u+v)^2}-\frac{1-uv}{(u-v)^2}-\frac{uv}{2}\right]\Psi^{(2)}_+(v),\\
-\left(\frac{m^2}{1-u^2}-\frac{M^2}{4}\right)\Psi^{(2)}_-(u)&=f_0\int_{-1}^1\frac{dv}{2\pi}\frac{1}{\sqrt{(1-u^2)(1-v^2)}}\left[\frac{1+uv}{(u+v)^2}-\frac{1-uv}{(u-v)^2}\right]\Psi^{(2)}_-(v),\\
\left(\frac{m^2}{1-u^2}-\frac{M^2}{4}\right)\Psi^{(1)}(u)&=f_0\int_{-1}^1\frac{dv}{2\pi}\frac{-2}{\sqrt{(1-u^2)(1-v^2)}}\left[-\frac{1-uv}{(u-v)^2}+\frac{1}{4}\right]\Psi^{(1)}(v),
\ea\ee 
with $\Psi^{(2)}_{\pm}(u)=\lim_{P\to\infty}\Psi^{(2)}_{P,\pm}(uP/2)$ and $\Psi^{(1)}(u)=\lim_{P\to\infty}\Psi^{(1)}_P(uP/2)$.
The string tension
$f_0=2\overline{\sigma}^2\lambda$, 
where $\overline{\sigma}\approx1.3578383|m|^{1/8}$.
For later convenience we introduce a dimensionless parameter 
\be
\gamma=\frac{f_0}{m^2}.
\ee

Now we introduce
\be
\Psi^{(1)}_{\pm}(u)=\frac{1}{2}\left(\Psi^{(1)}(u)\pm\Psi^{(1)}(-u)\right),
\ee
which satisfy $\Psi^{(2)}_{\pm}(u)=-\Psi^{(2)}_{\pm}(-u)$ and 
$\Psi^{(1)}_{\pm}(u)=\pm\Psi^{(1)}_{\pm}(-u)$.
By performing the change of variable $u=\tanh\theta$ and denoting $\Theta(\theta)=\Psi(u)$,
we obtain
\be
\ba 
&\left(m^2-\frac{M^2}{4\cosh^2\theta}\right)\Theta_+^{(2)}(\theta)=f_0\int_{-\infty}^{\infty}\frac{d\theta^{\prime}}{2\pi}\left(\frac{2\cosh(\theta-\theta^{\prime})}{\sinh^2(\theta-\theta^{\prime})}+\frac{\sinh\theta \sinh\theta^{\prime}}{2\cosh^2\theta\cosh^2\theta^{\prime}}\right)\Theta_+^{(2)}(\theta^{\prime}),\\
&\left(m^2-\frac{M^2}{4\cosh^2\theta}\right)\Theta^{(2)}_-(\theta)=f_0\int_{-\infty}^{\infty}\frac{d\theta^{\prime}}{2\pi} \frac{2\cosh(\theta-\theta^{\prime})}{\sinh^2(\theta-\theta^{\prime})}\Theta^{(2)}_-(\theta^{\prime}),\\
&\left(m^2-\frac{M^2}{4\cosh^2\theta}\right)\Theta^{(1)}_{\pm}(\theta)=f_0\int_{-\infty}^{\infty}\frac{d\theta^{\prime}}{2\pi}\left(\frac{2\cosh(\theta-\theta^{\prime})}{\sinh^2(\theta-\theta^{\prime})}-\frac{1}{2\cosh\theta\cosh\theta^{\prime}}\right)\Theta^{(1)}_{\pm}(\theta^{\prime}).
\label{eq:theta}
\ea\ee 
Notice that since $\Theta^{(1)}_-(\theta)=-\Theta^{(1)}_-(-\theta)$,
integration for the second term on the r.h.s. of $\Theta^{(1)}_-$ vanishes.
Therefore,
$\Theta^{(1)}_-$ and $\Theta^{(2)}_-$ take the same set of solutions. 
Eq.\eqref{eq:BS1} and \eqref{eq:theta} are the BS equations of the Ising ladder theory, which we shall analyze in detail below.

Now the problem reduces to solving the eigenvalue equation 
\be
\label{eq:HM}
\hat{H}\Theta=M^2\Theta, 
\ee
with 
\be 
\hat{H}\Theta(\theta)=4\cosh^2\theta\left[m^2\Theta(\theta)-f_0\int_{-\infty}^{\infty}G(\theta|\theta^{\prime})\Theta(\theta^{\prime})\frac{d\theta^{\prime}}{2\pi}\right].
\label{eq:eigs}
\ee 
Four sets of wavefunctions are classified through the fermion number parity ($\mc{N}$) on each chain and chain exchanging parity $U_{\text{ex}}$: 
\begin{align*}
\Theta_+^{(2)}:& \quad \mc{N}\text{ - even},\quad U_{\text{ex}}\text{ - even},\\\nonumber
\Theta_-^{(2)}:& \quad \mc{N}\text{ - even}, \quad U_{\text{ex}}\text{ - odd},\\\nonumber
\Theta_+^{(1)}:& \quad \mc{N}\text{ - odd},\quad U_{\text{ex}}\text{ - even},\\\nonumber
\Theta_-^{(1)}:& \quad \mc{N}\text{ - odd},\quad U_{\text{ex}}\text{ - odd}.
\end{align*}
The kernels $G(\theta|\theta^{\prime})$ for all cases can be read off from Eq.~\eqref{eq:theta}.

\subsection{Weak coupling expansion}
In the presence of weak interchain coupling that
$f_0\ll m^2$,
we expect the physics of weakly confined quarks.
As $\gamma\to 0$,
each eigenvalue $M_n^2$ of Eq.~\eqref{eq:eigs} approaches $4m^2$ from the above.
Following Eq.~\eqref{eq:lp}, 
low energy expansion of $M_n^2$ includes fractional powers of $\gamma$.
We introduce
\be
t=\gamma^{\frac{1}{3}},
\ee 
and expand $M$ in Eq.\eqref{eq:HM} as
\be
\frac{M^2}{4m^2}=1+zt^2+\sum_{k=3}^{\infty}c_kt^k.
\label{eq:mexpand}
\ee 
First,
we start from the approximated solutions of Eq.~\eqref{eq:theta}
\be\ba
&\Theta_{+,0}^{(2)}(\theta)=\Theta_{-,0}^{(2)}(\theta)=\zeta(\theta)-\zeta(-\theta), \quad\Theta^{(1)}_{-,0}(\theta)=\Theta^{(2)}_{-,0}(\theta),
\quad\Theta_{+,0}^{(1)}(\theta)=\zeta(\theta)+\zeta(-\theta),
\label{eq:Theta}
\ea\ee
with 
\be\ba
&\zeta(\theta)=\frac{1}{2}\int_{-\infty}^{\infty}\frac{e^{\frac{i}{\gamma}S(\beta)}}{\sinh(\theta+\beta-i0)}d\beta,\\
&S(\beta)=\frac{M^2}{4m^2}\tanh\beta-\beta.
\ea\ee
Error of the solution is captured by 
$[\hat{H}-M^2]\Theta_0(\theta)$, 
which can be calculated explicitly in powers of $t$.

\paragraph{\bf{Meson spectrum for $\Theta_{+}^{(2)}.$}} 
Here we introduce an error function $\delta^{(2)}_+(\theta)$ that captures the deviation between the two sides in Eq.~\eqref{eq:HM} when the trial solution for $\Theta_{+}^{(2)}$ (Eq.~\eqref{eq:Theta}) is inserted,
\be\ba
\delta^{(2)}_+(\theta)&\equiv\frac{1}{4m^2\sinh\theta}[\hat{H}-M^2]\Theta^{(2)}_+(\theta)\\
&=\int_{-\infty}^{\infty}e^{\frac{i}{\gamma}S(\beta)}\left[\frac{M^2}{4m^2}\frac{1}{\cosh\beta}+\frac{i\gamma}{4}\frac{\sinh\beta}{\cosh^2\beta}\right.\\
&+\left.\frac{\gamma}{\pi}\frac{\partial}{\partial\beta}\left(\frac{\cosh^2\theta}{\sinh\theta}\left(\frac{\theta+\beta}{\sinh(\theta+\beta)}-\frac{\theta-\beta}{\sinh(\theta-\beta)}\right)-\frac{\beta}{2\cosh\beta}\right)\right]d\beta.
\label{eq:err1}
\ea\ee 
The r.h.s. of Eq.~\eqref{eq:err1} can be expanded in $t$ utilizing Eq.~\eqref{eq:mexpand}.
At the leading order in $t$,
only the first term in the integrand contributes.
By introducing $u=-\beta/t$,
Eq.~\eqref{eq:err1} is approximated by
\be
\delta^{(2)}_+(\theta)\approx t\int_{-\infty}^{\infty}e^{\frac{iu^3}{3}-iuz}du=t\text{Ai}(-z).
\label{eq:lead1}
\ee 
Requiring $\delta^{(2)}_+(\theta)$ to vanish at this order leads to $z=z_n$ with Ai$(-z_n)=0$.
Higher order corrections can be found by requiring $\delta^{(2)}_+(\theta)$ to vanish order by order, 
which fixes $c_k$ in \eqref{eq:mexpand} order by order, more details are given in App.~\ref{app:weak}.
As a result, for the first several terms we obtained
\be 
\frac{(M_{+,n}^{(2)})^2}{4m^2}\approx1+z_nt^2+\frac{z_n^2}{5}t^4-\left(\frac{313}{420}+\frac{3z_n^3}{175}\right)t^6, 
\ee
with the meson mass $M_{+,n}^{(2)}$, $n=1,2,\cdots$ determined for Eq.~\eqref{eq:err1}.

\paragraph{Meson spectrum for $\Theta^{(2)}_-$.} 
The error function for $\Theta^{(2)}_-$ can be organized as
\be\ba
\delta^{(2)}_-(\theta)&\equiv\frac{1}{4m^2\sinh\theta}[\hat{H}-M^2]\Theta^{(2)}_-(\theta)\\
&=\int_{-\infty}^{\infty}e^{\frac{i}{\gamma}S(\beta)}\left[\frac{M^2}{4m^2}\frac{1}{\cosh\beta}
+\frac{\gamma}{\pi}\frac{\partial}{\partial\beta}\frac{\cosh^2\theta}{\sinh\theta}\left(\frac{\theta+\beta}{\sinh(\theta+\beta)}-\frac{\theta-\beta}{\sinh(\theta-\beta)}\right)\right]d\beta.
\label{eq:err2}
\ea\ee 
Performing the expansions we find the same leading order result for $\delta_-^{(2)}$ as that in Eq.~\eqref{eq:lead1}.
Results for the first several orders are obtained as
\be 
\frac{(M_{-,n}^{(2)})^2}{4m^2}\approx1+z_nt^2+\frac{z_n^2}{5}t^4-\left(\frac{52}{105}+\frac{3z_n^3}{175}\right)t^6. 
\ee

\paragraph{Meson spectrum for $\Theta^{(1)}_-$.}  $\Theta^{(1)}_-$ takes the same solutions as $\Theta_-^{(2)}$.

\paragraph{Meson spectrum for $\Theta^{(1)}_+$.} The error function for $\Theta^{(1)}_+$ reads
\be\ba
&\delta^{(1)}_+(\theta)=\frac{1}{4m^2\cosh\theta}[\hat{H}-M^2]\Theta^{(1)}_+(\theta)\\
=&\int_{-\infty}^{\infty}e^{\frac{i}{\gamma}S(\beta)}\left[\frac{M^2}{4m^2}\frac{-\sinh\beta}{\cosh^2\beta}+\frac{\gamma\cosh\theta}{\pi}\frac{\partial}{\partial\beta}\left(\frac{\theta+\beta}{\sinh(\theta+\beta)}+\frac{\theta-\beta}{\sinh(\theta-\beta)}\right)+\frac{\gamma}{2}\left(\frac{\beta}{\pi}-\frac{i}{2}\right)\right]d\beta.
\ea\ee
The leading order of $\delta^{(1)}_+(\theta)$ is given by 
\be
\delta^{(1)}_+(\theta)\approx t^2\int_{-\infty}^{\infty}ue^{\frac{iu^3}{3}-iuz}du,
\ee
such that $z=\tilde{z}_n$ is determined through Ai$^{\prime}(-\tilde{z}_n)=0$.
Combining with higher order results,
we have 
\begin{align}
\frac{(M^{(1)}_{+,n})^2}{4m^2}\approx &\,1+\tilde{z}_n t^2+\left(-\frac{11}{20 \tilde{z}_n} - \frac{2 \tilde{z}_n}{15} + \frac{\tilde{z}_n^2}{3}\right)t^4\\\nonumber
&+\left(\frac{19}{600} + \frac{121}{800 \tilde{z}_n^3} + \frac{11}{150 \tilde{z}_n} + \frac{2 \tilde{z}_n}{225} + \frac{649 \tilde{z}_n^2}{3150} - \frac{89 \tilde{z}_n^3}{450}\right)t^6.
\end{align}

\begin{figure}[htp]
    \centering
    \includegraphics[width=0.7\textwidth]{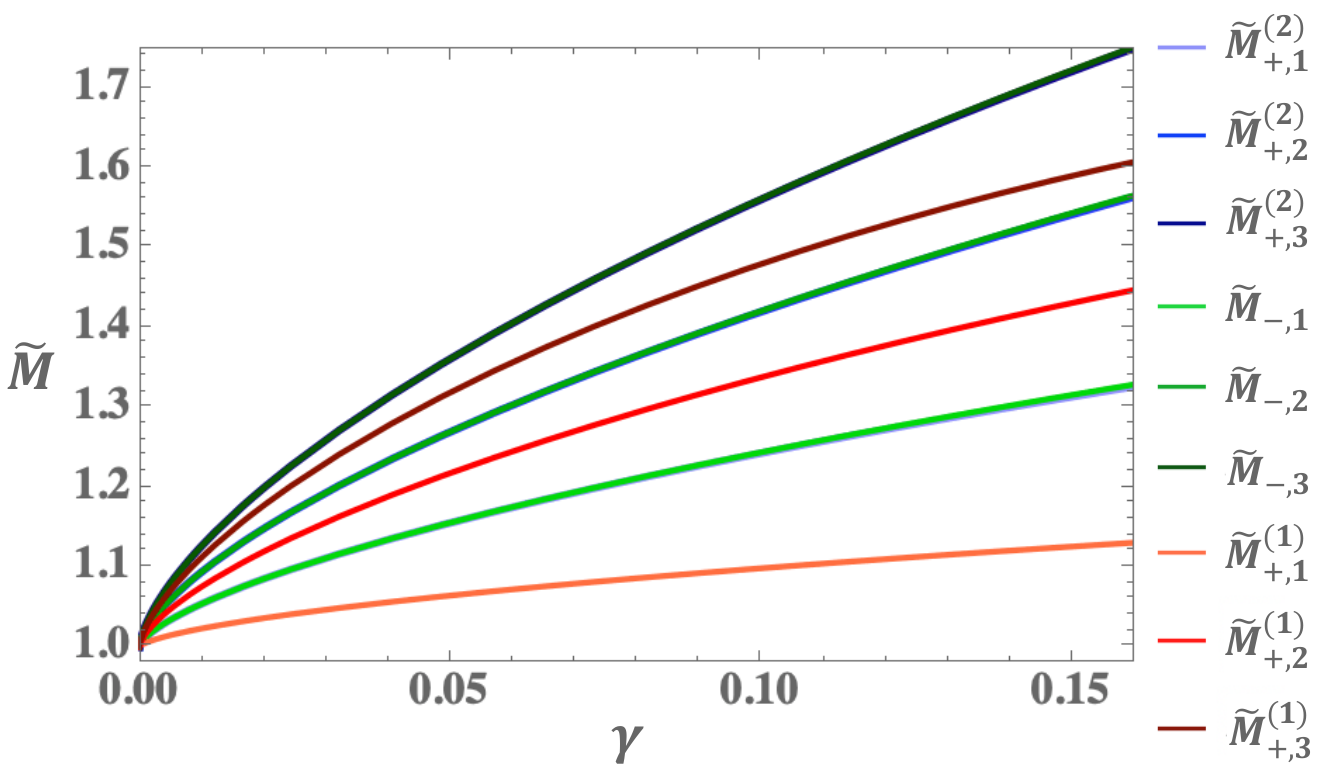}
    \caption{$\widetilde{M}\equiv M/(2m)$ v.s. $\gamma$ for lowest three energy levels in each sector in the weak coupling region.}
    \label{fig:weakexp}
\end{figure}

Fig.~\ref{fig:weakexp} shows the results for the first three levels in each sector.
In the weak coupling limit,
the nearly degenerate spectra of the $U_{\text{ex}}$-odd mesons and the $\mc{N}$- and $U_{\text{ex}}$-even mesons gradually separate as $\gamma$ increases.

\subsection{Numerical solutions}
For general coupling, the BS equations cannot be solved analytically and we resort to numerical solutions.
To avoid dealing with singular kernels in Eq.~\eqref{eq:theta},
we introduce the conjugate wavefunctions as follows,
\be\ba 
&\Theta^{(\cdot)}(\theta)=\int_{-\infty}^{\infty} d\nu\,\psi^{(\cdot)}(\nu)e^{-i\nu\theta},\\
&\psi^{(\cdot)}(\nu)=\int_{-\infty}^{\infty}\frac{d\theta}{2\pi}\Theta^{(\cdot)}(\theta)e^{i\nu\theta}.
\ea\ee 
Noticing that $\psi^{(2)}_{\pm}(\nu)=-\psi^{(2)}_{\pm}(-\nu)$ and $\psi^{(1)}_{\pm}(\nu)=\pm\psi^{(1)}_{\pm}(-\nu)$,
Fourier transform is performed on both sides of Eq.~\eqref{eq:theta} [{\it{cf}}. App.~\ref{app:FourierBS} for more details],
leading to
\be\ba 
&8\left(m^2+f_0\nu\tanh\frac{\pi\nu}{2}\right) \psi^{(2)}_+(\nu)-f_0\frac{\nu}{\cosh(\pi\nu/2)}\int_{-\infty}^{\infty}\frac{\nu^{\prime}}{\cosh(\pi\nu^{\prime}/2)}\psi^{(2)}_+(\nu^{\prime})d\nu^{\prime}\\
=&(M_+^{(2)})^2\int_{-\infty}^{\infty}d\nu^{\prime}\,[K(\nu-\nu^{\prime})-K(\nu+\nu^{\prime})]\psi^{(2)}_+(\nu^{\prime}),\\
&8\left(m^2+f_0\nu\tanh\frac{\pi\nu}{2}\right) \psi^{(2)}_-(\nu)
=(M_-^{(2)})^2\int_{-\infty}^{\infty}d\nu^{\prime}\,[K(\nu-\nu^{\prime})-K(\nu+\nu^{\prime})]\psi^{(2)}_-(\nu^{\prime}),\\
&8\left(m^2+f_0\nu\tanh\frac{\pi\nu}{2}\right) \psi^{(1)}_+(\nu)+\frac{f_0}{\cosh(\pi\nu/2)}\int_{-\infty}^{\infty}d\nu^{\prime}\frac{\psi^{(1)}_+(\nu^{\prime})}{\cosh(\pi\nu^{\prime}/2)}\\
=&(M_+^{(1)})^2\int_{-\infty}^{\infty}d\nu^{\prime}\,[K(\nu-\nu^{\prime})+K(\nu+\nu^{\prime})]\psi^{{(1)}}_+(\nu^{\prime}),\\
&8\left(m^2+f_0\nu\tanh\frac{\pi\nu}{2}\right) \psi^{(1)}_-(\nu)=(M_-^{(1)})^2\int_{-\infty}^{\infty}d\nu^{\prime}\,[K(\nu-\nu^{\prime})-K(\nu+\nu^{\prime})]\psi^{(1)}_-(\nu^{\prime}),
\label{eq:ft2}
\ea\ee 
with $K(\nu)=\nu/(2\sinh(\pi\nu/2))$
and all kernels being regular.
These equations can be directly solved by discretizing the 
integration.
Results for the first several levels are shown in Tab.~\ref{tab:ff}.

\begin{table}[h]
\vspace{2pt}
\centering 
\begin{tabular}{ccccccccc}
    \toprule 
    $\gamma$ & 0.001 &0.01 &0.1&1&10&100&1000&10000\\
    \midrule
    $(M_{+,1}^{(2)})^2/(4m^2)$ & 1.023&1.110&1.550&4.150&23.578&195.831&1847.731&18143.526\\
    $(M_{+,2}^{(2)})^2/(4m^2)$ & 1.042&1.197&2.024&7.619&56.084&520.744&5107.159&50782.707\\
    $(M_{+,3}^{(2)})^2/(4m^2)$ &1.062&1.269&2.446&10.952&87.856&836.802&8266.714&82379.560\\
    \midrule 
    $M_{-,1}^2/(4m^2)$ &1.023&1.111&1.552&4.220&24.992&214.114&2046.722&20188.009\\
    $M_{-,2}^2/(4m^2)$ &1.042&1.197&2.025&7.656&56.602&526.491&5166.371&51380.246\\
    $M_{-,3}^2/(4m^2)$ &1.062&1.269&2.447&10.976&88.173&840.192&8301.269&82727.076\\
    \midrule 
    $(M_{+,1}^{(1)})^2/(4m^2)$ &1.010&1.048&1.226&2.136&7.558&48.315&411.725&3905.063\\
    $(M_{+,2}^{(1)})^2/(4m^2)$ &1.033&1.155&1.791&5.905&40.275&364.798&3551.394&35232.269\\
    $(M_{+,3}^{(1)})^2/(4m^2)$ &1.051&1.2233&2.237&9.292&72.082&680.203&6702.398&66738.617\\
    \bottomrule
\end{tabular}
\caption{$M_n^2/(4m^2)$ for $n$-th meson in all the sectors obtained by numerically solving Eq.~\eqref{eq:ft2}.}
\label{tab:ff}
\end{table}

\begin{figure}[h]
    \centering
    \includegraphics[width=0.80\textwidth]{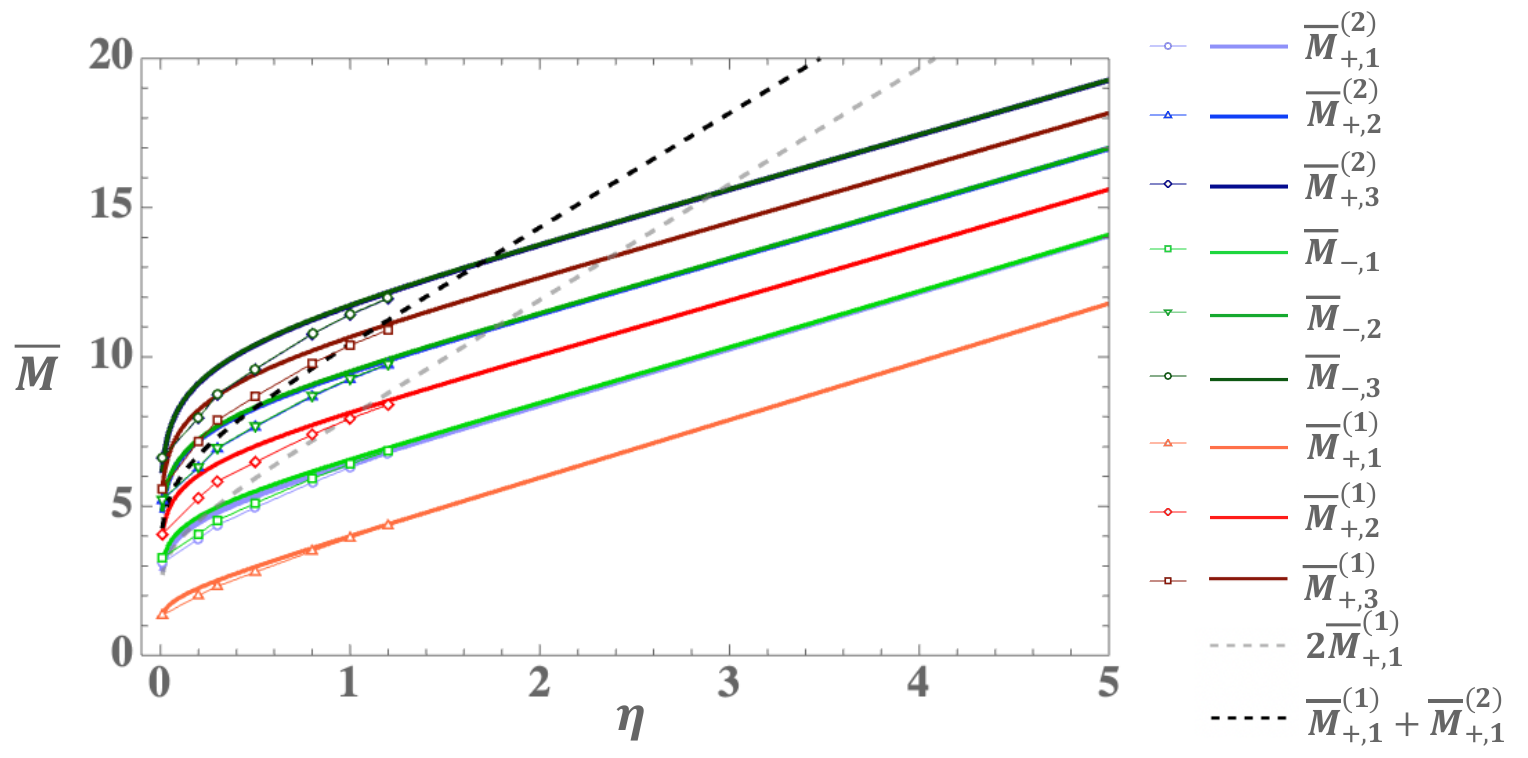}
    \caption{$\overline{M}(\eta)\equiv M(\eta)/\lambda^{4/7}$ v.s. $\eta$ with smooth curves solved from Eq.~\eqref{eq:ft2}. The joint lines with marks are solved from the same set of equations by replacing the bare $f_0$ with renormalized ones [Eq.~(\ref{eq:renormalizedf}) in Sec.~\ref{sec:tffsa}].
    The gray lines denote two kinds of two-particle thresholds [Sec.~\ref{sec:integrable}].}
    \label{fig:BSFF}
\end{figure}

We further introduce a dimensionless parameter 
\be
\eta=\frac{m}{\lambda^{4/7}}.
\label{eq:eta}
\ee 
Fig.~\ref{fig:BSFF} shows $M(\eta)/\lambda^{4/7}$ for several low energy states.
We find that the mass of the lightest interchain meson is always smaller than the intrachain one.
This can be understood intuitively from the number of interchain domain walls in the two cases.
In the former case,
the length can tend to vanish,
while in the latter,
it must maintain a finite length to be distinguished from the ground state.
Consistent with the weak coupling analysis,
the sets $M^{(2)}_{+}$ and $M_{-}$ tend to be degenerate as the coupling decreases.

The BS equation based on the two-particle projection works well except for $\eta\to 0$,
where the bare string tension $f_0\to0$.
However,
the quantum Ising ladder is still in the ordered phase in this limit,
leading to finite energy cost if we flip the spin in the ground state,
\emph{i.e.}, finite string tension is expected.
In the following section,
we will improve the result by using a renormalized tension.

\section{Beyond the two-particle space}
\label{sec:tffsa}
The BS equations describe the excitations constituted by two particles.
As mentioned in the previous section,
as $\eta\to0$ the bare string tension $f_0 \to 0$,
which does not respect the sustained confinement for the domain walls due to the presence of the order and interchain Ising coupling.
In this section,
we apply the truncated free fermionic space approach (TFFSA),
which includes general number of particles,
to further confirm the meson picture and quantitatively improve the BS results.

\subsection{The setup}
In this section,
we apply the TFFSA \cite{yurov1990,yurov1991,Fonseca2003}
which deals with Eq.~\eqref{eq:action} defined with finite spatial range $L$.
Consider massive fermions first.
Fermions in each chain split into two sectors:
the Neveu-Schwartz (NS) sector and the Ramond (R) sector.
The discrete momentum takes the value $2\pi k/R$,
with $k\in\mathbb{Z}+1/2$ for the NS sector and $k\in \mathbb{Z}$ for the R sector,
which is related to the rapidity $\theta_k$ through $mR\sinh\theta_k=2\pi k$.

Considering two chains,
we have four sectors: NS-NS, R-R, NS-R, and R-NS,
with the first two decoupled from the latter two.
The former two sectors are considered in the following,
as the latter ones yield the same physics.
To study bulk physics of Eq.~\eqref{eq:action} with open boundary
and focus on the low energy part,
we allow both even and odd fermion numbers in each chain, truncating them to a finite value while maintaining the same fermion parity for both chains.
The basis are chosen as 
\be
|k_1,\cdots,k_m\rangle_{1,\text{NS}}\otimes|l_1,\cdots,l_n\rangle_{2,\text{NS}},
\quad |k_1,\cdots,k_m\rangle_{1,\text{R}}\otimes|l_1,\cdots,l_n\rangle_{2,\text{R}},
\ee
with $m+n$ being even number.

Similar to Sec.~\ref{sec:BS},
since non-vanishing $F^{\sigma}(\theta_1\cdots\theta_n|\theta_1^{\prime}\cdots\theta_m^{\prime})=_{\text{NS}}\langle \theta_1,\cdots,\theta_n|\sigma(0)|\theta^\prime_1,\cdots,\theta^\prime_m\rangle_{\text{R}}$ requires even $m+n$,
Hilbert space with $\mc{N}$-even ($+$) decoupled from $\mc{N}$-odd ($-$) part.
Furthermore,
the basis can be organized into $U_{\text{ex}}$-even ($+$) and $U_{\text{ex}}$-odd ($-$) parts,
which are also decoupled from each other.
We use the eigenvalues of ($\mc{N}$, $U_{\text{ex}}$) to label different sectors of the Hilbert space.
The Hamiltonian can be block-diagonalized for the four sectors ($+,+$), ($+,-$), ($-,+$), ($-,-$),
after inserting the fermion energy and form factors for finite size system \cite{Fonseca2003} [{\it{cf}}. App.~\ref{app:TCSA}].
Since we are interested in the mass spectrum,
we can further restrict to the zero total-momentum subspace $\sum_{j}k_j+\sum_{i}l_i=0$.
Here we take the truncation by requiring $\sum_{j}|k_j|+\sum_{i}|l_i| < \Lambda_c$.

At $\eta=0$, \emph{i.e.}, 
starting from the massless fermions, 
the truncated conformal space approach (TCSA) is applied 
to the two coupled $c=1/2$ CFTs.
For each CFT,
a set of basis that interpolates the NS and R sector \cite{yurov1991} is applied.
Here we generalize the method \cite{yurov1991} by including odd-number fermion [App.~\ref{app:TCSA}].
The Hilbert space can be decomposed in the same way in the massive case.

\begin{figure}[htp]
    \centering
    \includegraphics[width=1\textwidth]{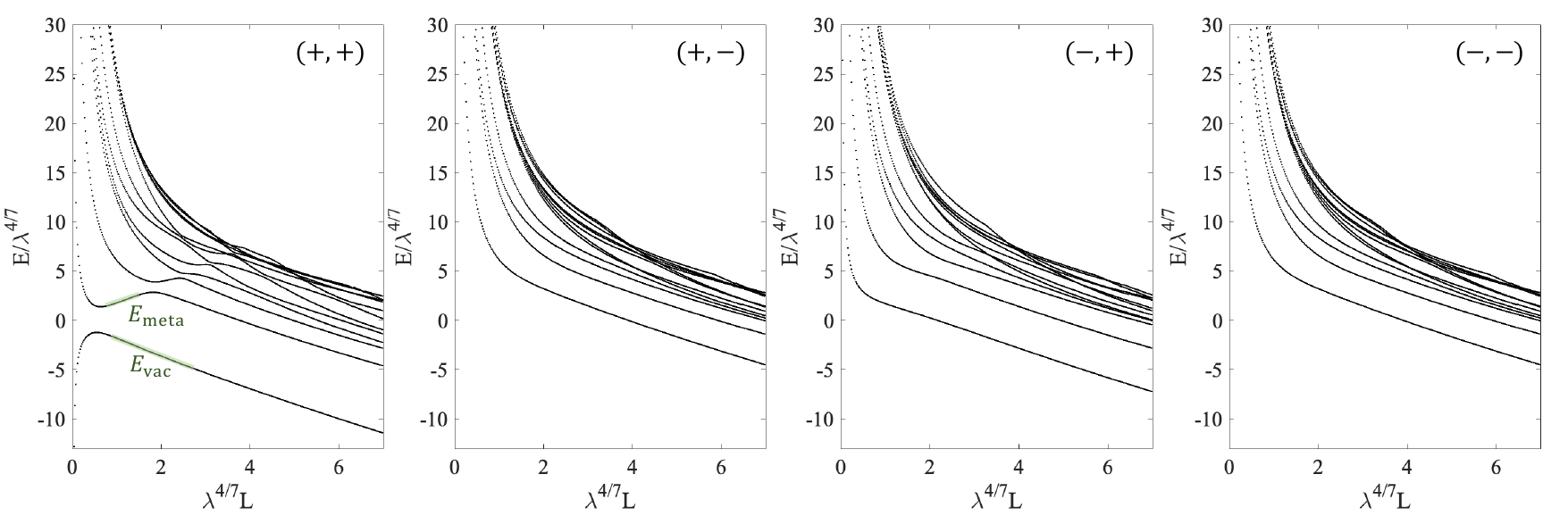}
    \caption{First several energy levels for each sector obtained from TFFSA for $\eta=1$ and level truncation at $ \Lambda_c = 12$.}
    \label{fig:eta1}
\end{figure}

\subsection{String tension}
As discussed previously,
the bare string tension $f_0$ cannot be directly applied to the limit $\eta\to0$.
Physically, this tension arises from interchain coupling and corresponds to the 
mean energy cost per unit length for propagating 
an intrachain domain wall, denoted as
$e_{\text{meta}}(m,\lambda)$.  To analyze this systematically, we introduce a meta-stable false vacuum
where
the interchain domain walls span the entire ladder without intrachain domain walls, and
its energy satisfies $E_{\text{meta}}(m,\lambda)=e_{\text{meta}}(m,\lambda)L$.
Consider the energy for the true vacuum $E_{\text{vac}}(m,\lambda)=e_{\text{vac}}(m,\lambda)L$
with $e_{\text{vac}}$ being the energy density of the true vacuum,
following the approach in \cite{zam2006}, the renormalized tension strength is then given by
\be
f_{\text{re}}(m,\lambda)=\Re [e_{\text{meta}}(m,\lambda)]-e_{\text{vac}}(m,\lambda),
\label{eq:renormalizedf}
\ee
with $\Re [\cdot]$ taking the real part of the quantity.

Relation between $E/\lambda^{4/7}$ and $L\lambda^{4/7}$ can be obtained for fixed $\eta$.
For example, 
results of the first several levels for $\eta=1$ is 
shown in Fig.~\ref{fig:eta1}.
Both the true vacuum and the false vacuum are found in the ($+,+$) sector, 
as illustrated in Fig.~\ref{fig:eta1}.
By fitting the slope, we obtain
\be
f_{\text{re}}(\eta\lambda^{4/7},\lambda)=\alpha(\eta)\lambda^{8/7}
\ee
with several $\alpha(\eta)$ listed in Tab.~\ref{tab:tension}.

By replacing $f_0\to f_{\text{re}}$ in Eq.~\eqref{eq:theta} and in the subsequent equations
we can obtain the mass spectrum from BS equatons with renormalized tension
[Fig.~\ref{fig:BSFF}].
The correction significantly modifies the meson mass behavior near $\eta=0$,
where $f_{\text{re}}$ approaches a finite constant in contrast to vanishing $f_0$.
The finite tension reflects the influence of interchain coupling and magnetic order at $\eta=0$, \textit{i.e.}, 
it costs energy for the intrachain domain wall to propagate.

\begin{table}[h]
    \caption{$\alpha$ obtained from TFFSA for a number of $\eta$s.}
      \vspace{2pt}
    \centering 
    \begin{tabular}{ccccccccc}
        \toprule 
        $\eta$ & 1.5 & 1.2 & 1 & 0.8 & 0.5 & 0.3 & 0.2 & 0\\
        \midrule
        $\alpha$  &3.890 & 3.611 &3.397 &3.081 & 2.574 & 2.223 & 1.902 & 1.329 \\
        \bottomrule
    \end{tabular}
    \label{tab:tension}   
\end{table}

\begin{table}[h]
    \centering 
    \begin{tabular}{cccccc}
        \toprule 
        $\eta$ & 0.3 & 0.5 & 1 & 2 & 5 \\
        \midrule
        $(+,+)$ & \makecell{4.04\\7.62} & \makecell{5.06\\8.07\\10.00} &
        \makecell{6.36\\8.96\\11.82}
        &\makecell{8.41\\11.38\\13.91} 
        & \makecell{14.03\\16.84\\18.93\\19.78} \\
        \midrule
        $(-,+)$ & \makecell{1.85\\6.18} & \makecell{2.65\\7.18} & \makecell{4.00\\8.22\\10.19} & \makecell{6.00\\10.08\\12.62} & \makecell{11.74\\15.50\\17.94\\20.10}\\
        \midrule
        $(\pm,-)$ &\makecell{4.58} & \makecell{5.60\\8.48} & \makecell{6.77\\9.68\\11.25} &\makecell{8.53\\11.49\\13.75} &\makecell{14.08\\16.95\\19.17\\20.97}\\
        \bottomrule
    \end{tabular}
    \caption{$M/\lambda^{4/7}$ of several lightest stable particle in each sector 
    identified from the TFFSA for a few $\eta$s.}
    \label{tab:tff}
\end{table}

\subsection{Stable particles and the degeneracy}
\begin{figure}[htp]
        \centering
        \includegraphics[width=1\textwidth]{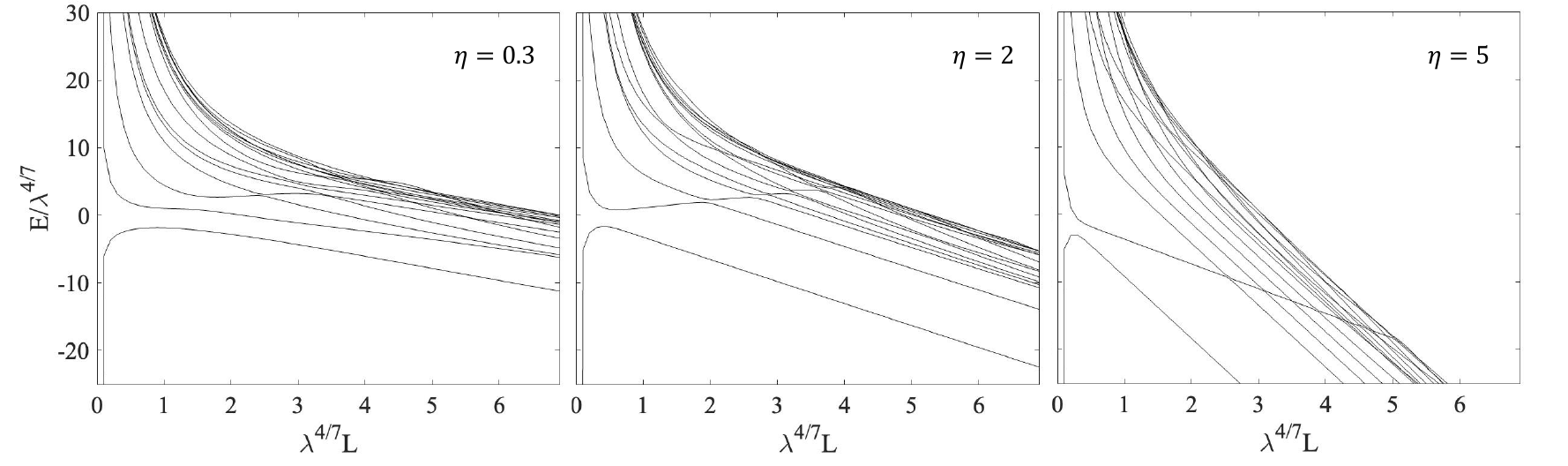}
        \caption{Energy spectra obtained by TFFSA in the $(+,+)$ sector with different $\eta$s.}
        \label{fig:evo}
    \end{figure}

The energy of a stable particle $A_i$ takes the form 
\be 
E_i(\eta,L) = e_{\text{vac}}(m,\lambda)L+M_i
\ee
for large $L$.
By identifying the lines parallel to the line of ground states,
the energy levels of stable particles can be found in all the four sectors,
with particle masses shown in Tab.~\ref{tab:tff}.
Fig.~\ref{fig:evo} shows the spectra in $(+,+)$ sector with several $\eta$'s as an illustration.
It can be observed that as $\eta$ grows,
number of stable particles increases.

$U_{\text{ex}}$-even and $U_{\text{ex}}$-odd lead to different spectra [Fig.~\ref{fig:eta1}], 
which is consistent with solutions of the BS equations,
indicating that the eigenfunctions split into the two sectors.
Moreover,
the ($+,-$) and ($-,-$) sectors give the same energy levels as we 
take the same truncation,
confirming the degeneracy found in the BS equation,
which goes beyond the two-particle approximation.

Intuitively,
the degeneracy that solely appears in the $U_{\text{ex}}$-odd sector
can be qualitatively understood using the Ising-limit picture.
Two chains with the same configuration, \textit{e.g.,}, $|\psi\rangle = |K_{c_1}\cdots K_{c_n}\rangle_1\otimes|K_{c_1}\cdots K_{c_n}\rangle_2$,
with $K_{j}$ representing a kink at site $j$,
belong to the $U_{\text{ex}}$-even sector.
$n$ takes even and odd number for states in $(+,+)$ and $(-,+)$, respectively,
whose energies differ in general.
While in the presence of interchain domain-walls,
we can
consider a $(-,-)$ state with component $(1-U_{\text{ex}})|\psi\rangle$,
where 
$|\psi\rangle = |K_{a_1}\cdots K_{a_{2n-1}}\rangle_1\otimes|K_{b_1}\cdots K_{b_{2m-1}}\rangle_2$.
Moving one kink to the other chain, \textit{e.g.,}
$|\tilde{\psi}\rangle = |K_{a_2}\cdots K_{a_{2n-1}}\rangle_1\otimes|K_{a_1}K_{b_1}\cdots K_{b_{2m-1}}\rangle_2$,
leading to the $(+,-)$-state $(1-U_{\text{ex}})|\tilde{\psi}\rangle$.
$a_1$ is assumed to be different from elements in $\{b_n\}$.
With both interchain and intrachain domain-wall size unchanged,
$|\psi\rangle$ and $|\tilde{\psi}\rangle$ have the same energy,
rendering $U_{\text{ex}}$-odd sectors degenerate.

\section{Integrable point and its vicinity}
\label{sec:integrable}
Similar to the Ising field theory, at specific value 
$\eta=0$ $(m=0)$ of the Ising ladder system [Eq.~\eqref{eq:action}] 
it becomes integrable for low-energy physics, 
known as the Ising$_h^2$ integrable field theory \cite{coupleCFT}.
This theory has been proposed to be realized in quantum magnet 
CoNb$_2$O$_6$ \cite{Xning} and has undergone detailed study 
in \cite{dark}, where exotic quasiparticles such as `dark particles' were identified.
Two key questions arise. 
The first one is what is the fermion or spin context of the `dark particles' 
corresponding to the breathers with odd parity in the sine-Gordon model \cite{dark}.
The second one is whether the `dark particles' remain stable after moving away from the integrability point.
This section will address both questions through systematic analysis of the model's fermionic representation 
and its perturbations away from integrability.
\subsection{Ising$_h^2$ integrable field theory}
At $\eta=0$,
the low-energy physics of the quantum Ising ladder is described by the Ising$_h^2$ integrable field theory (IIFT).
Using the Majorana spinors $\psi^{(a)}=(\psi^{(a)}_R,\psi^{(a)}_L)^\intercal$ [Eq.~\eqref{eq:phi}],
the Dirac spinor can be constructed as
\be
\chi=\frac{1}{\sqrt{2}}(\psi^{(1)}+i\psi^{(2)})\equiv(\chi_R,\chi_L)^\intercal.
\ee
In terms of a massless bosonic field $\phi(x,t)=\phi_R(x-t)+\phi_L(x+t)$
with $\phi_{R}$ and $\phi_{L}$ being the right- and left-moving waves,
the bosonization can be performed through \cite{2isingboson,banks}
\be
\chi_R=\frac{1}{\sqrt{N}}\alpha_R:e^{i\phi_R}:,\quad
\chi_L=\frac{1}{\sqrt{N}}\alpha_L:e^{-i\phi_L}:,
\label{bosonization}
\ee
where $\alpha_R$ and $\alpha_L$ are the Klein factors \cite{2isingboson} 
that guarantee $\{\chi_R,\chi_L\}=0$. Following Eq.~(\ref{bosonization}),
when the interchain coupling vanishes,
the $H^{(a)}_{\text{Ising}}(\tau=0,0)$ [Eq.~\eqref{eq:free}] 
for the two chains
can be cast into a free massless boson theory. When the 
interchain coupling is present, Eq.~(\ref{eq:action}) at 
$\tau=0$ becomes \cite{coupleCFT}
\be
H^b_{\text{IIFT}}=\frac{1}{4\pi}\int dx \left[\frac{1}{2}\left(\frac{\partial\phi}{\partial t}\right)^2+\frac{1}{2}\left(\frac{\partial\phi}{\partial x}\right)^2-\xi\cos\left(\hat{\beta}\phi(x)\right)\right],\quad \hat{\beta}=\frac{1}{2}
\label{eq:isingh2}
\ee
with the application of the following bosonization rules,
\be\ba
\sigma^{(1)}\sigma^{(2)}\sim :\cos\phi/2:, \quad&
\mu^{(1)}\mu^{(2)}\sim:\sin\phi/2:,\\
(\epsilon^{(1)}+\epsilon^{(2)})/2\sim:\cos\phi:, \quad&
(\epsilon^{(1)}-\epsilon^{(2)})/2\sim:\cos\tilde{\phi}:,
\ea\ee
where $\mu^{(a)}(x)$ is the continuum limit of $\mu_j^{(a)}$ [Eq.~\eqref{eq:op}], 
and $\tilde{\phi}$ is the dual field of $\phi$ satisfying 
$\partial\phi/\partial t=-\partial\tilde{\phi}/\partial x$.
In addition, $\phi$ is defined on the $\mathbb{Z}_2$ orbifold with $\phi$ identified as $-\phi$ \cite{GINSPARG1988153}.
The rescaled interchain coupling $\xi=\sqrt{2}\lambda$ \cite{BASEILHAC2001607}.

In the convention of sine-Gordon theory described by the Hamiltonian $H_{\text{SG}}=\int dx(\partial_{\mu} \phi)^2/2+(4\pi)^{-1}\xi\cos(\beta\phi)$,
Eq.~\eqref{eq:isingh2} takes the same form as a reflectionless model with $\beta^2/(8\pi)=1/8$ (referred to as SG$_{1/8}$).
Eq. \eqref{eq:isingh2} is characterized by the $\mathcal{D}_8^{(1)}$ Lie algebra \cite{coupleCFT}, accommodating
8 types of particles:
six breathers $B_{n}\,(n=1,\cdots,6)$ and two solitons ($A_{\pm}$).
The mass relations are
\be 
m^0_{B_n}=m^0_{B_1}\sin(n\pi/14)/\sin(\pi/14), \quad
m^0_{A_{\pm}}=m^0_{B_1}/(2\sin(\pi/14)).
\label{eq:mass}
\ee

Global properties of the IIFT follows the well-studied SG$_{1/8}$ model\cite{direct,bookBosonization}.
Here we further expose their significant consequences for the corresponding microscopic spin model.
In the SG$_{1/8}$,
parity conjugation $\mc{C}$ that operates as $\mc{C}\phi\mc{C}^{-1} = -\phi$ leads to 
$\mc{C}|B_n\rangle=(-1)^n|B_n\rangle$ and $\mc{C}|A_{\pm 1}\rangle = |A_{\mp1}\rangle$.
$A_{\pm}$ carries topological charge $\pm1$, and
$e^{\pm i\tilde{\phi}}$ is referred to as $\pm1$-charge changing operator.
As a result, both $\cos\phi$ and $\cos\tilde{\phi}$,
related to local transverse spin operation, 
cannot connect the $\mc{C}$-even states with $\mc{C}$-odd states.
As a generalization, 
Ref.~\cite{dark} shows that $B_{1,3,5}$, termed as dark particles, cannot be directly excited from the ground state ($\mc{C}$-even) through any (quasi-)local spin operations.
In other words,
once $B_1$ is prepared, 
it remains stable against spontaneous decay.

\begin{figure}[htp]
    \centering
    \includegraphics[width=1\textwidth]{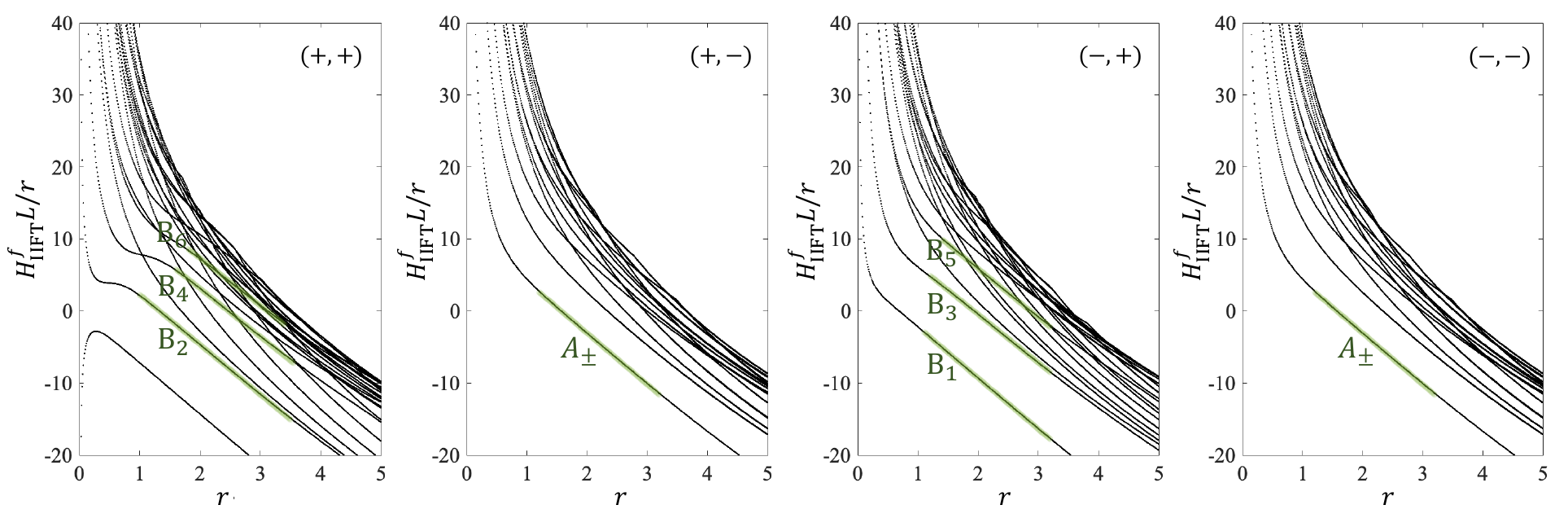}
    \caption{First several energy levels for each sector obtained from TCSA with level truncation $\Lambda_c = 14$. 
    The green lines illustrate the particle energy levels predicted by the IIFT.}
    \label{fig:cftfit}
\end{figure}
\subsection{Mass spectrum obtained from TCSA}
The bosonized theory offers us global properties of the emergent particles.
However, the physical content of the these particles remains unclear.
At the integrable point,
we can perform the TCSA to determine the mass relation between stable particles
which can be further classified according to
$\mc{N}$ and $U_{\text{ex}}$ parities.
Similar to the massive case,
we have four sectors according to $\mc{N}$ and $U_{\text{ex}}$ parities, leading to the energy levels shown in Fig.~\ref{fig:cftfit}.
Breathers $B_2$, $B_4$ and $B_6$ are identified in the $(+,+)$ sector,
while the $B_1$, $B_3$ and $B_5$ are in the $(-,+)$ sector.
Solitons $A_{\pm}$ live in the $U_{\text{ex}}$-odd sector.
By introducing the dimensionless scale \cite{yurov1991}
\be
r=\frac{(2\pi \lambda)^{4/7}}{2\pi}L,
\ee
we can determine the relation between eigenvalues of $H_{\text{IIFT}}L/r$ and $r$.
The mass ratios are obtained in Tab.~\ref{tab:cft},
which is in good agreement with Eq.~\eqref{eq:mass}. An even better agreement may be achieved by exploiting renormalization group improvements of the TCSA method \cite{Feverati:2006ni,Lencses:2015bpa}, which we leave for the future work.

\begin{table}[h]
    \caption{Mass ratios calculated from TCSA and its comparison with IIFT results}
      \vspace{3pt}
    \centering 
    \begin{tabular}{ccccccc}
        \toprule 
        &$m_{B_2}/m_{B_1}$ & $m_{B_3}/m_{B_1}$ & $m_{B_4}/m_{B_1}$ & $m_{B_5}/m_{B_1}$ &$m_{B_6}/m_{B_1}$&$m_{A_{\pm}}/m_{B_1}$\\
        \midrule
        TCSA&1.88792 & 2.6568 &3.37866 & 3.97082 &4.19243&2.29063\\
        IIFT&1.94986 & 2.80194 & 3.51352 & 4.04892 & 4.38129 & 2.24698\\
        \bottomrule
    \end{tabular}
    \label{tab:cft}
\end{table}

\begin{figure}[htp]
    \centering
    \includegraphics[width=0.8\textwidth]{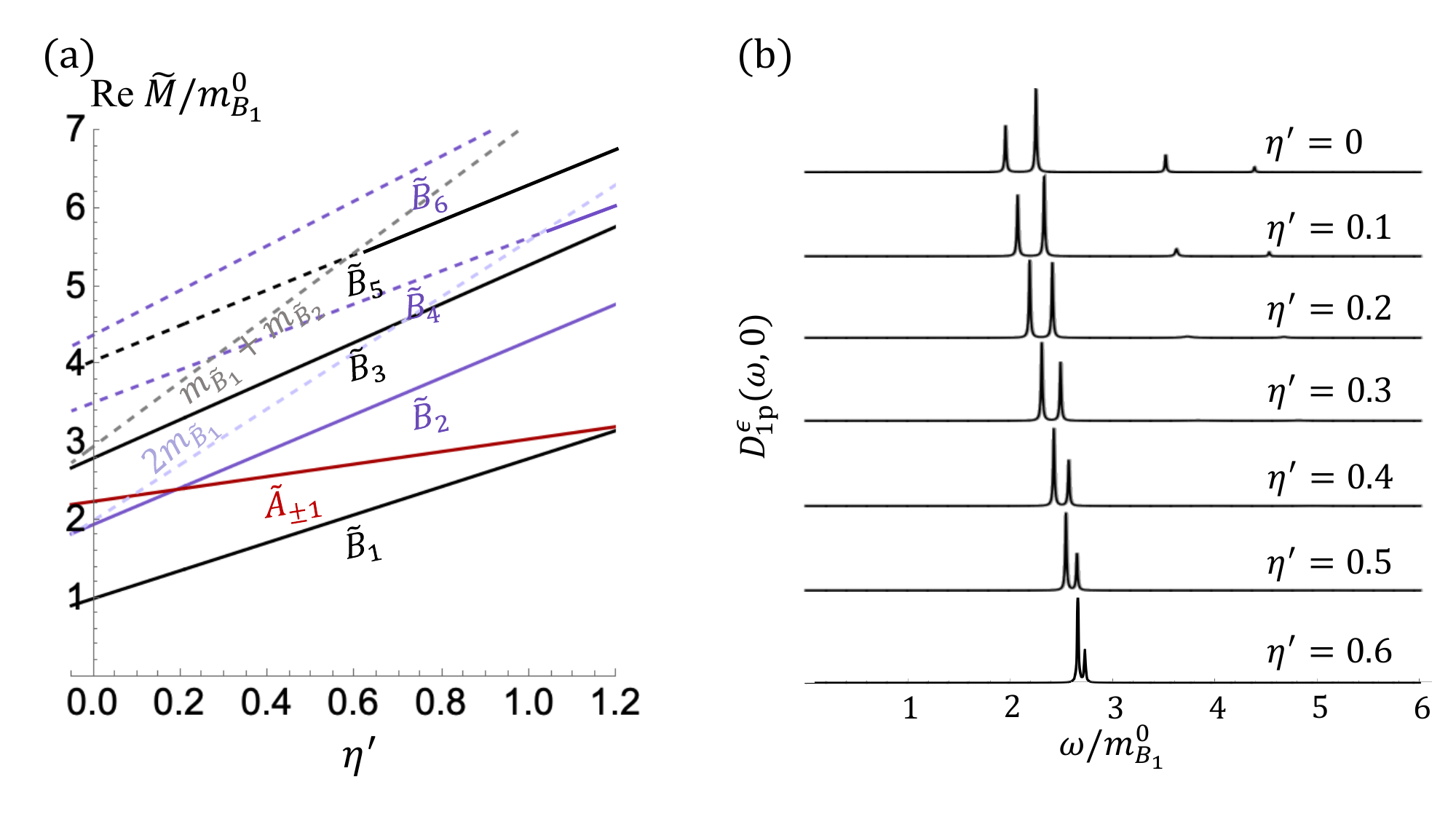}
    \caption{(a) Relation between real part of particle mass and $\eta^{\prime}$ obtained from first order 
    form factor perturbation theory. 
    The black lines represent $\tilde{B}_{1,3,5}$ and the purple lines represent $\tilde{B}_{2,4,6}$,
    where the solid segments indicate stable particle and the dashed segments indicate unstable ones.
    The dashed lines with light purple color denotes $2m_{\tilde{B}_1}$ threshold, above which the $\tilde{B}_{2,4,6}$ would decay. And the dashed gray line denote $m_{\tilde{B}_1}+m_{\tilde{B}_2}$ threshold, above which $\tilde{B}_{3,5}$ would decay.
    (b) Transverse spin DSF contributed by single particle channels, with the peak evolution shown for different $\eta^\prime$s.
    Each single-particle peak is broadened for better illustration.}
    \label{fig:FFPT}
\end{figure}
\subsection{Spectrum and particle stability away from the integrable point}
By slightly tuning the transverse field away from $g=1$,
the model immediately becomes non-integrable.
The perturbed Hamiltonian in the bosonized form is obtained as 
\be
H_{\text{pert}} = H_{\text{IIFT}}+ \rho\int dx~ \cos\phi(x),
\ee 
with $\rho$ being the rescaled perturbation strength $g_{\text{1d,c}}-g$,
which appears similarly as a double-frequency sine-Gordon theory \cite{DELFINO1998}.
For the unperturbed IIFT,
form factors $
F^{\hat{\mathcal{O}}}_{P_1,\dots,P_n}(\theta_1>\theta_2>\dots>\theta_n)=\langle 0|\hat{\mathcal{O}}|P_1(\theta_1)P_2(\theta_2)\dots P_n(\theta_n)\rangle$ of a local observable $\hat{\mathcal{O}}$ can 
be derived \cite{dark,KAROWSKI1978455, Smirnov,Lukyanov1997},
where the $j$-th particle of type $P_j$ carries rapidity $\theta_j$.
The unperturbed asymptotic in-state (the ket) is an eigenstate of Eq.\eqref{eq:isingh2} with 
eigenenergy $E_n=\sum_{j=1}^n m_{P_j} \cosh\theta_j$ and eigen momentum
$Q_n=\sum_{j=1}^n m_{P_{j}} \sinh\theta_j$ \cite{Mussardobook}.
In the following, 
the perturbed quantities are labelled by tilt.

Form factor perturbation theory \cite{DELFINO1996,DELFINO2006}
provides a framework to study the effects of perturbations on integrable 
quantum field theories.
By requiring the normalization of the vacuum and single-particle states after perturbation,
the leading order corrections to the mass square of the 
$r^{th}$ particle $m_{r}^2$ follow
\begin{subequations} \begin{align}
&\delta \text{Re} m_r^2\simeq 2\rho F_{rr}^{\cos\phi}(i\pi,0),\label{eq:re}\\
&\text{Im} m_r^2=-\sum_{a\leq b,m_a+m_b\leq m_r}m_c\Gamma_{c\to ab}\simeq -\rho^2\sum_{a\leq b,m_a+m_b\leq m_r}2^{1-\delta_{a,b}}\frac{|F_{rab}^{\cos\phi}(i\pi,\theta_a,\theta_b)|^2}{m_rm_a|\sinh\theta_a|}\label{eq:im},
\end{align} \end{subequations}  
with $\theta_a$ and $\theta_b$ determined from energy and momentum conservation through the decay process $P_r \to P_aP_b$.
The vacuum energy correction $\delta E_{\text{vac}}\simeq \rho \langle0|\cos\phi|0\rangle$.
$\Gamma_{a\to cb}$ denotes the decay width of $P_c$ through the channel $P_c \to P_aP_b$.
Form factor coefficients $F_{rr}^{\cos\phi}(i\pi,0)/\langle \cos\phi\rangle$ are obtained as 
$-5.63525,~-14.4129,~-21.7686,~-23.3134,~-28.9486,\\-39.4467~,-11.2705$ 
for $r=B_{1,\cdots, 6}$ and $A_+,A_-$.
Also,
\begin{align}
|F_{B_4B_1B_1}^{\cos\phi}|^2/\langle\cos\phi\rangle^2=69.8245\,,\\\nonumber
|F_{B_5B_1B_2}^{\cos\phi}|^2/\langle\cos\phi\rangle^2=8.66299\,,\\\nonumber
|F_{B_6B_1B_1}^{\cos\phi}|^2/\langle\cos\phi\rangle^2=3.38827\,,\\\nonumber
|F_{B_6B_2B_2}^{\cos\phi}|^2/\langle\cos\phi\rangle^2=24.2841\,.
\end{align}
Using known results \cite{BASEILHAC2001607}
\be \ba
&\langle\cos\phi\rangle_{\rho=0}=2.278284275...\lambda^{4/7},\\
&m_{B_1}^0=2.2716402481...\lambda^{4/7},
\ea \ee
and introducing a dimensionless parameter
\be
\eta^{\prime} = \frac{2\pi\rho}{\lambda^{4/7}},
\ee  
we can determine the particle mass [Fig.~\ref{fig:FFPT} (a)].
Eq.~\eqref{eq:im} implies multiparticle thresholds of stability.
As shown in Fig.~\ref{fig:FFPT}$(a)$,
breathers heavier than $2m_{B_1}$,
namely $\widetilde{B}_{n\geq3}$,
immediately become unstable as $\rho$ turns on.
Although $\widetilde{m}_{A_{\pm 1}}>2\widetilde{m}_{B_1}$,
there's no decay channel from $A_{\pm1}$ to lighter particles, 
and thus $\widetilde{A}_{\pm 1}$ survive.
By tuning $g<g_{\text{1d,c}}$,
corresponding to $m>0$ in previous sections,
For $\widetilde{B}_{1,2,3}$ and $\widetilde{A}_{\pm 1}$ each acquires a heavier mass without decay.
The masses of $\widetilde{A}_{\pm1}$ and $\widetilde{B}_2$ gradually touch each other as $\eta^\prime$ increases.
Moreover,
continuously increasing $\eta^\prime$  leads to a sequence of reformation of stable $\tilde{B}_5$, $\tilde{B}_4$ and $\tilde{B}_6$.
While for $g>g_{\text{1d,c}}$,
$\widetilde{B}_2$ survives only in a small region and decays upon reaching $2\widetilde{m}_{B_1}$,
where a second-order phase transition has been predicted when $\widetilde{m}_{B_1} = 0$ \cite{BAJNOK2001503}.

The mass thresholds for particle decay can also be determined from the BS equation [Fig.~\ref{fig:BSFF}],
where the masses $m_{\tilde{B}_{1/3/5}}\to M_{+,1/2/3}^{(1)}$,
$m_{\tilde{B}_{2/4/6}} \to M_{+,1/2/3}^{(2)}$ and $m_{\tilde{A}_{\pm}}\to M_{-,1}$ near $\eta=0$.
Consistent results for the stability and energy evolutions of these particles can be observed.

\subsection{The spin dynamics}
To investigate the previously mentioned dark property of $\tilde{B}_1$ in a general context,
as well as its experimental relevance,
the spin dynamical structure factor (DSF) is considered.
Taking advantage of the analytical form factors,
we consider the transverse spin ($\epsilon^{(1,2)}$) DSF,
corresponding to $(\cos\phi+\cos\tilde{\phi})/2$ in the bosonized form.
The DSF for operator $\mc{O}$ is obtained through the Fourier transformation of the correlation function as 
\be
D^{\mc{O}}(\omega,q)=\int_{-\infty}^{\infty}dx dt\langle\mc{O}(x,t)\mc{O}(0,0)\rangle e^{i\omega t}e^{-iqx},
\ee
with transferred energy $\omega$ and momentum $q$.

We consider the parameter region 
where the stability sustained for the lightest three particles,
and focus on the single particle channels that dominate the spectra.
Following Fermi Second Golden Rule, single particle spectra contribution to the zero transfer momentum 
DSF follows 
\be \ba
D^{\epsilon}_{1p}(\omega,0)=\sum_{\{P_r\}}\frac{|\langle \widetilde{0}|\cos\phi+\cos\tilde{\phi}|\widetilde{P}_r\rangle|^2}{m_{P_r}^{0}}W_{P_r},
\ea \ee 
with $W_{P_r}=\delta(\omega-\widetilde{m}_{P^{\prime}_r})$ for a stable particle $\tilde{P}_r$,
and $W_{P_r}=\Gamma_{P_r^{\prime}}/((\omega-\widetilde{m}_{P^{\prime}_r})^2+\Gamma_{P^{\prime}_r}^2)$
for an unstable $\tilde{P}^{\prime}_r$.
$\Gamma_{P^{\prime}_r}$ is the summation of decay width from all the decay channels of $\tilde{P}^{\prime}_r$.
The Jaccobian $d\widetilde{Q}_r/d\theta_r|_{\theta_r=0}=m_{P_r}^0+\mathcal{O}(\rho^2)$ is inserted for the perturbed states,
by noticing that momentum is conserved through the uniform perturbation.
The transition $\langle \widetilde{0}|\cos\phi+\cos\tilde{\phi}|\widetilde{P}_r\rangle$ is calculated with finite volume regularization \cite{POZSGAY2008,marton2019},
where we generalize the perturbative expansion of the ground state to single particle states and obtain the DSF (see App.~\ref{app:FFPT}).

As a result,
$\widetilde{B}_1$ involves $\mc{C}$-odd states
while the vacuum includes $\mc{C}$-even states after perturbation.
$|\widetilde{0}\rangle\to \widetilde{B}_1$ is still forbidden from (quasi-)local operations, 
signaled by the missing contribution at $\omega = \widetilde{m}_{B_1}$ shown in Fig.~\ref{fig:FFPT}$(b)$.
Fig.~\ref{fig:FFPT}$(b)$ illustrates the decay of $B_{4,6}$ in the DSF.
It also shows the relative spectral weight between $\tilde{B}_2$ and $\tilde{A}_{\pm1}$ tends to increase when $\eta$ grows.

\section{Conclusions}
In this article,
we study emergent mesons in the ordered region of a quantum Ising ladder.
We show that the meson physics originates 
from the confinement of free fermions in each TFIC of the ladder.
Taking advantage of fermion number parity for each chain and chain exchanging parity,
we obtain four decoupled BS equations,
each of which characterizes a set of mesons with the corresponding property.
These equations are studied by projecting onto the two-particle subspace,
with the infinite momentum frame applied to make the projection valid.
Analytical approximate solutions are obtained in the weak coupling regime,
and numerical calculations are performed for general coupling case.
Degeneracy between the $\mc{N}$-odd and even sectors within the $U_{\text{ex}}$-odd subspace has been found,
which tend to be nearly degenerate with the $\mc{N}$- and $U_{\text{ex}}$-even sector as the interchain coupling decreases.
The lightest meson formed in the $\mc{N}$-odd and $U_{\text{ex}}$-even sector is always smaller than the other three cases,
which can be understood as possessing the shortest interchain domain-wall length.
The bare BS equations work well except in the $\eta\to0$ limit,
where a renormalization of the string tension is required.
We obtain the $f_{\text{re}}$ by numerically fitting the energy density for the false vacuum and ground state through TFFSA.
Simultaneously,
the degeneracy aforementioned is further confirmed through this approach,
which is beyond the two-particle approximation.

Such characterization is extended to the $\eta=0$ case,
where the Ising$_h^2$ integrable model had been raised
and generally studied in a bosonic form.
We identify that the breathers and solitons in the bosonic 
framework are involved in $U_{\text{ex}}$-even and $U_{\text{ex}}$-odd sectors in the fermionic, or spin representations.
The odd breathers $B_{1,3,5}$ and even ones $B_{2,4,6}$ are found within the $\mc{N}$-odd and even subspace, respectively.
Evolution of the particles when tuning away from the integrable point is discussed.
It is shown that $B_{1,2,3}$ and $A_{\pm}$ remain stable,
whereas others decay into the continuum once $\eta\neq0$.
Tuning $\eta$ to larger values leads to more stable particles,
with $B_5$, $B_4$, $B_6$ reappearing sequentially.
Stable particles, in addition to those inherited from Ising$_h^2$ ones,
are also expected as $\eta$ increases,
with the particle number tending to infinity in the Ising limit.

For the TFFSA approach applied at $\eta=0$,
we generalize the CFT basis from Ref.~\cite{yurov1991} to the odd-fermion-number situation.
It is necessary to consider the spin model with open boundary
to obtain all the particles predicted by the Ising$_h^2$ integrable theory,
where both odd and even fermion number configurations are allowed.

It is worth noticing that the Hamiltonian Eq.~\eqref{eq:lattice} contains simple interactions,
making it possible for a direct quantum simulations of this model,
\textit{e.g.}, via Rydberg arrays \cite{rydberg}.
It has also been proposed that the model can be effectively realized in quantum magnet CoNb$_2$O$_6$ \cite{Xning}.
This work provides a physical picture of the excitations in these systems, 
serving as a potential guidance for further manipulations of these states by leveraging the parity properties introduced here.
On the other hand, 
the two-particle confinement framework can have a general application to a class of coupled bi-partite systems in which
there are well-defined quasi-particles or well characterized wavefunctions for each system, such as 
bilayer graphene, Heisenberg ladder model and \emph{etc}.

\section{Acknowledgments}
We thank Aliosha Litvinov, Misha Alfimov, G{\'{a}}bor Tak{\'{a}}cs and Rong Yu for helpful discussions.
The work of J.W. is supported by the National Natural Science Foundation of China Grant Nos. 12450004, 12274288
and the Innovation Program for Quantum Science and Technology Grant No. 2021ZD0301900. The work of Y.J. is partly supported by Startup Funding no. 3207022217A1 of Southeast University and by the NSF of China through Grant No. 12247103.

\appendix
\section{The BS kernels}
\label{app:BSkernel}
Here we give explicit forms of the BS kernels. 
The asymptotic states are denoted by
\be
c_p^{\dagger}c_q^{\dagger}|0\rangle=|p,q\rangle, \qquad
\langle0|c_qc_p=\langle q,p|,
\ee
where $c^{\dagger}$ and $c$ are the fermionic creation and annihilation operators and $p,q$ are the corresponding momenta.
In terms of rapidity $\theta$,
we also introduce the fermion operators
\be
c(\theta)=\sqrt{\omega(p)}c_p,\qquad c^{\dagger}(\theta)=\sqrt{\omega(p)}c_p^\dagger,
\ee 
where $p=m\sinh\theta$ and $\omega(p)=\sqrt{p^2+m^2}$.
Under this convention,
the projection to two-particle subspace in the main text leads to
\be 
\langle p_1,p_2|\int\frac{dq_1}{2\pi}\int\frac{dq_2}{2\pi}\Psi^{(2,0)}(q_1,q_2)|q_1,q_2\rangle=\Psi^{(2,0)}(p_2,p_1)=-\Psi^{(2,0)}(p_1,p_2).
\ee

The Ising form factor in the infinite volume is given by \cite{Berg_1979,BABELON1992113}
\be\ba
F^{\sigma}(\theta_1,\cdots,\theta_k|\theta^\prime_1,\cdots,\theta^\prime_l)=&\langle\theta_1,\cdots,\theta_k|\sigma(0)|\theta_1^\prime,\cdots,\theta_l^\prime\rangle\\
=&i^{\left[\frac{k+l}{2}\right]}\overline{\sigma}\prod_{0<i<j\leq k}\tanh\left(\frac{\theta_i-\theta_j}{2}\right)\prod_{0<p<q\leq l}\tanh\left(\frac{\theta^\prime_p-\theta^\prime_q}{2}\right)\\
&\times\prod_{0<s<m,0<t\leq m}\coth\left(\frac{\theta_s-\theta_t^\prime}{2}\right)\\
=&\prod_{i=1}^k\sqrt{\omega(p_i)}\prod_{j=1}^l\sqrt{\omega(q_j)}\langle p_1,\cdots p_k|\sigma(0)|q_1,\cdots,q_l\rangle.
\ea\ee
with $\overline{\sigma}=(1.357834...)|m|^{1/8}$ and $[\cdots]$ denoting the floor function.

\section{Weak coupling expansion}
\label{app:weak}
In this section, we present detailed calculation of the error functions and the meson masses in the weak coupling expansion.
\subsection{Airy function and its derivative}
We first review some basics of Airy function. 
The Airy function is defined as 
\be 
\text{Ai}(-z)=\frac{1}{2\pi}\int_{-\infty}^{\infty}e^{\frac{iu^3}{3}-iuz}du.
\ee 
For later convenience, we introduce
\be
F_k(-z)=\frac{1}{2\pi}\int_{-\infty}^{\infty}(iu)^ke^{\frac{iu^3}{3}-iuz}du,
\ee 
which satisfies
\be 
(k+1)F_k(-z)=F_{k+3}(-z)+zF_{k+1}(-z).
\label{eq:recursion}
\ee 
The zeros of the Airy function and its derivative will be particularly useful for us. 
Therefore we present some of their properties.
\paragraph{Zeros of Airy function}
Let us denote the solutions of Ai$(-z)=0$ by $z_n$ ($n=1,2,\ldots$).
For these values,
$F_0(-z_n)=0$ implies $F_2(-z_n)=0$ due to Eq.~\eqref{eq:recursion}.
It then follows from Eq.~\eqref{eq:recursion} that
\be
F_3(-z_n)=-z_nF_1(-z_n),\quad
F_4(-z_n)=2F_1(-z_n),\quad \,...
\ee
It is clear that all $F_{k}(-z_n)$ are proportional to $F_1(-z_n)$. We can introduce the following quantity
\begin{align}
I_k\equiv \frac{F_{k}(-z_n)}{F_1(-z_n)},\qquad k=0,1,2,\ldots
\end{align}
They can be computed straightforwardly. The values of the first few $I_k$'s read
\begin{align}
\label{eq:Iks}
&I_0=0,& &I_1=1,& &I_2=0, & &I_3=-z_n\\\nonumber
&I_4=2,& &I_5=z_n^2,& & I_6=-6z_n, & &I_7=-z_n^3+10,\\\nonumber
&I_8=12z_n^2,& &I_9=z_n^4-52z_n,& & I_{10}=-20z_n^3+80, & &\cdots
\end{align}
\paragraph{Zeros of derivative of Airy function}
We denote the solutions of Ai$^{\prime}(-{z})=0$ by $\tilde{z}_n$ ($n=1,2,\ldots$).
For these values,
$F_2(-\tilde{z}_n)=-\tilde{z}F_0(-\tilde{z}_n)$ and $F_3(-\tilde{z}_n)=F_0(-\tilde{z}_n)$.
All $F_k(-\tilde{z}_n)$ are proportional to $F_0(-\tilde{z})$.
Similarly, we define
\begin{align}
J_k\equiv \frac{F_k(-\tilde{z})}{F_0(-\tilde{z})},\qquad k=0,1,2,\ldots
\end{align}
The values of the first few $I_k$'s read
\begin{align}
\label{eq:Jks}
&J_0=1,& &J_1=0,& &J_2=-\tilde{z}_n, & &J_3=1\\\nonumber
&J_4=\tilde{z}_n^2,& &J_5=-4\tilde{z}_n,& & J_6=4-\tilde{z}_n^4, & &J_7=9\tilde{z}_n^2,\\\nonumber
&J_8=\tilde{z}_n^3-28\tilde{z}_n,& &J_9=-9\tilde{z}_n^3-7\tilde{z}_n^2+28,& & J_{10}=-\tilde{z}_n^4+100\tilde{z}_n^2, & &\cdots
\end{align}

\paragraph{Numerical values} The numerical values of the first few $z_n$ and $\tilde{z}_n$ are given by following table

\begin{table}[h]
    \centering 
    \begin{tabular}{c|cccc}
        \toprule 
        $n$ & $1$ & $2$ & $3$ & $4$ \\
        \midrule
        $z_n$ & 2.33811 & 4.08795 &5.52056 & 6.78671\\
        $\tilde{z}_n$ &1.01879 & 3.24820 & 4.82010 & 6.16331 \\
        \bottomrule
    \end{tabular}
    \caption{First several solutions for Ai$(-z)=0$ and Ai$^\prime(-\tilde{z})=0$.}
    \label{tab:zn}
\end{table}

\subsection{Perturbative calculation of meson masses}
\label{app:theta2p}
We denote
\be \ba
&[\hat{H}-M^2]\Theta^{(2)}_{\pm,0}(\theta)=4m^2\cosh^2\theta[\hat{\Omega}\Theta^{(2)}_{\pm,0}(\theta)+\hat{G}^{(2)}_{\pm}\Theta^{(2)}_{\pm,0}(\theta)],\\
&[\hat{H}-M^2]\Theta^{(1)}_{+,0}(\theta)=4m^2\cosh^2\theta[\hat{\Omega}\Theta_{+,0}^{(1)}(\theta)+\hat{G}^{(1)}_+\Theta^{(1)}_{+,0}(\theta)],\\
\ea \ee 
where 
\be
\hat{\Omega}\Theta(\theta)=\left(1-\frac{M^2}{4m^2}\frac{1}{\cosh^2\theta}\right)\Theta(\theta)=-\frac{\partial S(\theta)}{\partial\theta}\Theta(\theta).
\ee
\subsubsection*{1) $\Theta^{(2)}_+$}
Following Eq.~\eqref{eq:theta},
we have 
\be\ba
\hat{G}^{(2)}_+\Theta^{(2)}_{+,0}(\theta)=-\gamma\int_{-\infty}^{\infty}G_0(\theta-\theta^{\prime})\Theta_{+,0}^{(2)}(\theta^{\prime})\frac{d\theta^{\prime}}{2\pi}+\frac{\gamma}{2}\frac{\sinh\theta}{\cosh^2\theta}\overline{\Theta}_{+,0}^{(2)},
\label{eq:gt}
\ea\ee
with 
\be
G_0(\theta)=\frac{2\cosh\theta}{\sinh^2\theta},\qquad 
\overline{\Theta}_{+,0}^{(2)}=-\int_{-\infty}^{\infty}\frac{\sinh\theta}{\cosh^2\theta}\Theta_{+,0}^{(2)}(\theta)\frac{d\theta}{2\pi}.
\ee 
In terms of $\zeta(\theta)$,
\be\ba
&-\gamma\int_{-\infty}^{\infty}G_0(\theta-\theta^{\prime})\zeta(\theta^{\prime})\frac{d\theta^{\prime}}{2\pi}\\
=&\frac{\partial}{\partial \theta}\int_{-\infty}^{\infty}\frac{\gamma}{\sinh(\theta-\theta^{\prime})}\frac{d\theta^{\prime}}{2\pi}\int_{-\infty}^{\infty}\frac{e^{\frac{i}{\gamma}S(\beta)}}{\sinh(\theta^{\prime}+\beta-i0)}d\beta\\
=&\int_{-\infty}^{\infty} \gamma e^{\frac{i}{\gamma}S(\beta)} d\beta\frac{\partial}{\partial\theta}\left\{\frac{i}{2}\frac{1}{\sinh(\theta+\beta-i0)}+\int_{-\infty}^{\infty}\frac{1}{\sinh(\theta-\alpha+i\pi/2)\sinh(\alpha-i\pi/2+\beta)}d\alpha\right\}\\
=&\frac{i}{2}\int_{-\infty}^{\infty}\gamma e^{\frac{i}{\gamma}S(\beta)}\frac{\partial}{\partial\beta}\left(\frac{1}{\sinh(\theta+\beta-i0)}\right)d\beta+\frac{\partial}{\partial\theta}\int_{-\infty}^{\infty}\gamma e^{\frac{i}{\gamma}S(\beta)}\frac{2(\beta+\theta)}{\sinh(\beta+\theta)}\frac{d\beta}{2\pi}\\
=&\frac{1}{2}\int_{-\infty}^{\infty}\frac{\partial S(\beta)}{\partial\beta}\frac{ e^{\frac{i}{\gamma}S(\beta)}}{\sinh(\theta+\beta-i0)}d\beta+\frac{\gamma}{\pi}\int_{-\infty}^{\infty}\frac{\partial}{\partial\beta}\left(\frac{\beta+\theta}{\sinh(\beta+\theta)}\right)e^{\frac{i}{\gamma}S(\beta)}d\beta.
\label{eq:g0}
\ea\ee
The second term in Eq.~\eqref{eq:gt} involves the following quantity
\be\ba
Y&\equiv-\frac{1}{2}\int_{-\infty}^{\infty}\frac{\sinh\theta}{\cosh^2\theta}\frac{d\theta}{2\pi}\int_{-\infty}^{\infty}\frac{e^{\frac{i}{\gamma}S(\beta)}}{\sinh(\theta+\beta+i0)}d\beta\\
&=\int_{-\infty}^{\infty}\left[\frac{i}{4}\frac{\sinh\beta}{\cosh^2\beta}-\frac{1}{2\pi}\frac{\partial}{\partial\beta}\frac{\beta}{\cosh\beta}\right]e^{\frac{i}{\gamma}S(\beta)}d\beta.
\ea\ee 
Such that,
\be\ba
&[\hat{\Omega}+\hat{G}_+^{(2)}]\zeta(\theta)\\
=&\frac{1}{2}\int_{-\infty}^{\infty}\frac{\Omega(\theta)-\Omega(\beta)}{\sinh(\theta+\beta-i0)}e^{\frac{i}{\gamma}S(\beta)}d\beta+\frac{\gamma}{\pi}\int_{-\infty}^{\infty}\frac{\partial}
{\partial\beta}\left(\frac{\beta+\theta}{\sinh(\beta+\theta)}\right)e^{\frac{i}{\gamma}S(\beta)}d\beta+
\frac{\gamma}{2}\frac{\sinh\theta}{\cosh^2\theta}Y\\
=&\frac{1}{2}\int_{-\infty}^{\infty}\frac{\sinh(\theta-\beta)}{\cosh^2\theta\cosh^2\beta}e^{\frac{i}{\gamma}S(\beta)}d\beta+\frac{\gamma}{\pi}\int_{-\infty}^{\infty}\frac{\partial}{\partial\beta}
\left(\frac{\beta+\theta}{\sinh(\beta+\theta)}\right)e^{\frac{i}{\gamma}S(\beta)}d\beta+\frac{\gamma}{2}\frac{\sinh\theta}{\cosh^2\theta}Y.
\ea\ee 
Combined with $\zeta(-\theta)$ part, we obtain [Eq.~(\ref{eq:Theta})]
\be\ba
[\hat{\Omega}+\hat{G}_+^{(2)}]\Theta^{(2)}_{+,0}(\theta)&=\int_{-\infty}^{\infty}e^{\frac{i}{\gamma}S(\beta)}\left[\frac{M^2}{4m^2}\frac{\sinh\theta}{\cosh^2\theta\cosh\beta}\right.\\
&+\left.\frac{\gamma}{\pi}\frac{\partial}{\partial\beta}\left(\frac{\theta+\beta}{\sinh(\theta+\beta)}-\frac{\theta-\beta}{\sinh(\theta-\beta)}-\frac{\sinh\theta}{2\cosh^2\theta}\frac{i\pi/2+\beta}{\cosh\beta}\right)\right]d\beta,
\label{eq:even}
\ea\ee
which gives $\delta_+^{(2)}$ in the main text.

The leading order correction to $(M^{(2)}_{+,n})^2/(4m^2)$ is given by $z=z_n$. We now explain how to obtain higher order corrections.
\begin{itemize}
\item Order $t^2$ \\ Expanding the error function at this order and requiring
\begin{align}
-it^2\int_{-\infty}^{\infty}du\,e^{\frac{iu^3}{3}-z_nu}c_3u=0
\end{align}
implies that
\begin{align}
c_3=0.
\end{align}
\item Order $t^3$\\
We require
\begin{align}
t^3\int_{-\infty}^{\infty}du\,e^{\frac{iu^3}{3}-z_nu}\left(z_n-\frac{u^2}{2}-ic_4u-\frac{2u^5i}{15}+\frac{u^3z_ni}{3}-\frac{c_3^2u^2}{2}\right)=0\,.
\end{align}
This leads to
\begin{align}
c_4=\frac{I_2}{2}-\frac{2I_5}{15}+\frac{z_nI_3}{3i^2}
\end{align}
where we have used the fact that $c_3=0$. Plugging in the explicit values of $I_k$'s in \eqref{eq:Iks}, we obtain
\begin{align}
c_4=\frac{z_n^2}{5}\,.
\end{align}
\end{itemize}
A similar procedure leads to
\begin{align}
c_5=0,\qquad c_6=-\frac{313}{420}-\frac{3z_n^3}{175}
\end{align}

\subsubsection*{2) $\Theta^{(2)}_-$}
For $\Theta^{(2)}_-$ we have
\be\ba
\hat{G}^{(2)}_{-}\Theta^{(2)}_{-,0}(\theta)=-\gamma\int_{-\infty}^{\infty}G_0(\theta-\theta^{\prime})\Theta^{(2)}_{-,0}(\theta^{\prime})\frac{d\theta^{\prime}}{2\pi},
\ea\ee
and 
\be
[\hat{\Omega}+\hat{G}^{(2)}_-]\Theta^{(2)}_{-,0}(\theta)=\int_{-\infty}^{\infty}e^{\frac{i}{\gamma}S(\beta)}\left[\frac{M^2}{4m^2}\frac{\sinh\theta}{\cosh^2\theta\cosh\beta}+\frac{\gamma}{\pi}\frac{\partial}{\partial\beta}\left(\frac{\theta+\beta}{\sinh(\theta+\beta)}-\frac{\theta-\beta}{\sinh(\theta-\beta)}\right)\right]d\beta.
\ee

The first five terms of $(M_{-,n}^{(2)})^2/(4m^2)$ are the same as those obtained for the $\Theta^+$ case. 
For the sixth order, we have
\be\ba
&t^5\int_{-\infty}^{\infty}du\,e^{\frac{iu^3}{3}-z_nu}\left(-ic_6u+\frac{5u^4}{24}+\frac{38iu^7}{315}-\frac{2u^{10}}{225}-\frac{u^2z_n}{2}-\frac{13iu^5z_n}{30}+\frac{2u^8z_n}{45}\right.\\
&\quad\quad\quad\quad\left.+\frac{z^2}{5}+\frac{iu^3z_n^2}{2}-\frac{37u^6z_n^2}{450}-\frac{iuz_n^3}{5}+\frac{u^4z_n^3}{15}-\frac{u^2z_n^4}{50}\right)=0
\ea\ee
which leads to
\begin{align}
 c_6^-=-\frac{52}{105}-\frac{3z_n^3}{175}
\end{align}

\subsubsection*{3) $\Theta^{(1)}_+$}
In this case,
\be\ba
&\hat{G}^{(1)}_+\Theta^{(1)}_{+,0}(\theta)=-\gamma\int_{-\infty}^{\infty}G_0(\theta-\theta^{\prime})\Theta_{+,0}^{(1)}(\theta^{\prime})\frac{d\theta^{\prime}}{2\pi}+\frac{\gamma}{2\cosh\theta}\overline{\Theta}^{(1)}_{+,0}(\theta),\\
&\overline{\Theta}^{(1)}_{+,0}(\theta)=\int_{-\infty}^{\infty}\frac{\Theta_{+,0}^{(1)}(\theta)}{\cosh\theta}\frac{d\theta}{2\pi}=\int_{-\infty}^{\infty}\frac{d\theta}{2\pi}\int_{-\infty}^{\infty}d\beta
\frac{e^{\frac{iS(\beta)}{\gamma}}}{\cosh\theta}
\left(\frac{1}{2\sinh(\theta+\beta-i0)}+\frac{1}{2\sinh(-\theta+\beta-i0)}\right)\\
&\qquad\quad=\int_{-\infty}^{\infty}e^{\frac{iS(\beta)}{\gamma}} \left(\frac{\beta}{\pi}-\frac{i}{2}\right)d\beta. 
\ea\ee
Using the result of Eq.~\eqref{eq:g0},
\be\ba
&[\hat{\Omega}+\hat{G}^{(1)}_+]\Theta^{(1)}_{+,0}(\theta)\\
=&\int_{-\infty}^{\infty}e^{\frac{i}{\gamma}S(\beta)}\left[\frac{M^2}{4m^2}\frac{-\sinh\beta}{\cosh\theta\cosh^2\beta}+\frac{\gamma}{\pi}\frac{\partial}{\partial\beta}\left(\frac{\theta+\beta}{\sinh(\theta+\beta)}+\frac{\theta-\beta}{\sinh(\theta-\beta)}\right)+\frac{\gamma}{2\cosh\theta}\left(\frac{\beta}{\pi}-\frac{i}{2}\right)\right]d\beta.
\ea\ee
The higher order corrections to the meson masses are obtained as follows
\begin{itemize}
\item Order $t^3$,
\begin{align}
t^3\int_{-\infty}^{\infty}du\,e^{\frac{iu^3}{3}-\tilde{z}_nu}(-i)u^2c_3=0,
\end{align}
which leads to
\begin{align}
c_3=0\,.
\end{align}
\item Order $t^4$,
\begin{align}
t^4\int_{-\infty}^{\infty}du\,e^{\frac{iu^3}{3}-\tilde{z}_nu}\left[-\frac{i}{4}-\frac{5u^3}{6}+u\tilde{z}_n-u^2i\left(c_4+\frac{2u^4}{15}-\frac{u^2\tilde{z}_n}{3}\right)\right]=0,
\end{align}
which leads to
\begin{align}
c_4=\frac{1}{J_2}\left(\frac{1}{4}+\frac{5J_3}{6}-\frac{zJ_4}{3}-\frac{2J_6}{15}\right)
\end{align}
Plugging in the explicit values of $J_k$'s in \eqref{eq:Jks}, we obtain
\begin{align}
c_4=-\frac{11}{20 \tilde{z}_n} - \frac{2 \tilde{z}_n}{15} + \frac{\tilde{z}_n^2}{3}\,.
\end{align}
\end{itemize}

Similar procedure gives
\begin{align}
c_5=0,\qquad 
c_6=\frac{19}{600} + \frac{121}{800 \tilde{z}_n^3} + \frac{11}{150 \tilde{z}_n} + \frac{2 \tilde{z}_n}{225} + \frac{649 \tilde{z}_n^2}{3150} - \frac{89 \tilde{z}_n^3}{450}\,.
\end{align}

\section{Fourier transform of the BS equations}
\label{app:FourierBS}
In this section we give the Fourier transformation of the BS equations. For the left-hand side of the BS equations \eqref{eq:theta}, we have
\be\ba 
&-\frac{M^2}{4}\int_{-\infty}^{\infty}\frac{d\theta}{2\pi}\frac{\Theta(\theta)}{\cosh^2\theta} e^{i\nu\theta}=-\frac{M^2}{4}\int_{-\infty}^{\infty}\frac{d\theta}{2\pi}\frac{e^{i\nu\theta}}{\cosh^2\theta}\left(\int_{-\infty}^{\infty}d\nu^{\prime}\,\psi(\nu^{\prime})e^{-i\nu^{\prime}\theta}\right)\\
&\qquad\quad\qquad\qquad\qquad\qquad=-\frac{M^2}{4}\int_{-\infty}^{\infty}\psi(\nu^{\prime})K(\nu-\nu^{\prime})d\nu^{\prime},
\ea\ee
where
\begin{align}
K(\nu)\equiv\frac{\nu}{2\sinh(\pi\nu/2)}\,.
\end{align}
The r.h.s. of the BS equations have two kinds of terms. Fourier transformation of the first kind reads
\be\ba
&\int_{-\infty}^{\infty}\frac{d\theta}{2\pi}e^{i\nu\theta}\int_{-\infty}^{\infty}\frac{d\theta^{\prime}}{2\pi}\frac{\cosh(\theta-\theta^{\prime})}{\sinh^2(\theta-\theta^{\prime})}\Theta(\theta^{\prime})\\
=&2\int_{-\infty}^{\infty}\frac{dx}{2\pi}\int_{-\infty}^{\infty}\frac{dy}{2\pi}\frac{\cosh(2x)}{\sinh^2(2x)}\int_{-\infty}^{\infty}d\nu^{\prime}\,\psi(\nu^{\prime})e^{-i(\nu^{\prime}-\nu)y}e^{i(\nu^{\prime}+\nu)x}\\
=&\int_{-\infty}^{\infty}\frac{dx}{2\pi}\frac{\cosh x}{\sinh^2 x}e^{i\nu x}\psi(\nu)
=-\frac{\nu}{2}\tanh\frac{\pi\nu}{2}\psi(\nu).
\ea\ee 
Fourier transformation of the second kind gives
\be\ba
&\int_{-\infty}^{\infty}\frac{d\theta}{2\pi}e^{i\nu\theta}\int_{-\infty}^{\infty}\frac{d\theta^{\prime}}{2\pi}\frac{\sinh\theta\sinh\theta^{\prime}}{\cosh^2\theta\cosh^2\theta^{\prime}}\Theta^{(2)}_+(\theta^{\prime})\\
=&\int_{-\infty}^{\infty}d\nu^{\prime}\,\psi^{(2)}_+(\nu^{\prime})\int_{-\infty}^{\infty}\frac{d\theta^{\prime}}{2\pi}\frac{\sinh\theta^{\prime}}{\cosh^2\theta^{\prime}}e^{-i\nu^{\prime}\theta^{\prime}}\left(\int_{-\infty}^{\infty}\frac{d\theta}{2\pi}\frac{\sinh\theta}{\cosh^2\theta}e^{i\nu\theta}\right)\\
=&\frac{1}{4}\frac{\nu}{\cosh(\pi\nu/2)}\int_{-\infty}^{\infty}d\nu^{\prime}\psi^{(2)}_+(\nu^{\prime})\frac{\nu^{\prime}}{\cosh(\pi\nu^{\prime}/2)}.
\ea\ee 
Fourier transform of the last kind of term reads
\be\ba
\int_{-\infty}^{\infty}\frac{d\theta}{2\pi}\int_{-\infty}^{\infty}\frac{d\theta^{\prime}}{2\pi}\,\frac{e^{i\nu\theta}\Theta^{(1)}_{\pm}
(\theta^{\prime})}{\cosh\theta\cosh\theta^{\prime}}=\frac{1}{4}\frac{1}{\cosh(\pi\nu/2)}\int_{-\infty}^{\infty}d\nu^{\prime}\frac{\psi^{(1)}_\pm(\nu^{\prime})}{\cosh(\pi\nu^{\prime}/2)}.
\ea\ee
As a result, the Fourier transformed BS equations follow by
\be\ba 
&m^2\psi^{(2)}_+(\nu)-\frac{M^2}{4}\int_{-\infty}^{\infty}d\nu^{\prime}K(\nu-\nu^{\prime})\psi^{(2)}_+(\nu^{\prime})\\
&\,\,=-f_0\nu\tanh\frac{\pi\nu}{2}\psi^{(2)}_+(\nu)+\frac{f_0\nu}{8\cosh(\pi\nu/2)}\int_{-\infty}^{\infty}d\nu^{\prime}\frac{\nu^{\prime}}{\cosh(\pi\nu^{\prime}/2)}\psi^{(2)}_+(\nu^{\prime}),\\
&m^2\psi^{(2)}_-(\nu)-\frac{M^2}{4}\int_{-\infty}^{\infty}d\nu^{\prime}K(\nu-\nu^{\prime})\psi^{(2)}_-(\nu^{\prime})=-f_0\nu\tanh\frac{\pi\nu}{2}\psi^{(2)}_-(\nu),\\
&m^2\psi^{(1)}_\pm(\nu)-\frac{M^2}{4}\int_{-\infty}^{\infty}d\nu^{\prime}K(\nu-\nu^{\prime})\psi^{(1)}_\pm(\nu^{\prime})\\
&\,\,=-f_0\nu\tanh\frac{\pi\nu}{2}\psi^{(1)}_\pm(\nu)-\frac{f_0}{8\cosh(\pi\nu/2)}\int_{-\infty}^{\infty}d\nu^{\prime}\frac{\psi^{(1)}_\pm(\nu^{\prime})}{\cosh(\pi\nu^{\prime}/2)}\,.
\label{eq:ft1}
\ea\ee

\section{Physical quantities in the finite size system}
\label{app:finite}
In this appendix, we present the formulas for physical quantities in finite volume used in the TFFSA calculations,
following \cite{Fonseca2003}.
For finite-size systems,
the form factors read
\be\ba
&_{NS,L}\langle \theta_1,\cdots,\theta_n|\sigma(0)|\theta^\prime_1,\cdots,\theta^\prime_m\rangle_{R,L}=S(L)\prod_{j=1}^n\tilde{g}(\theta_{j})\prod_{i=1}^mg(\theta_{i})F^{\sigma}(\theta_1,\cdots,\theta_n|\theta^\prime_1,\cdots,\theta^\prime_m),
\ea\ee
with
\be\ba
&S(L)=\exp\left\{\frac{(mL)^2}{2}\int\int_{-\infty}^{\infty}\frac{d\theta_1d\theta_2}{(2
\pi)^2}\frac{\sinh\theta_1\sinh\theta_2}{\sinh(mL\cosh\theta_1)\sinh(mL\cosh\theta_2)}\log\left|\coth\frac{\theta_1-\theta_2}{2}\right|\right\},\\
&g(\theta)=e^{\kappa(\theta)}/\sqrt{mL\cosh\theta},\quad
\tilde{g}(\theta)=e^{-\kappa(\theta)}/\sqrt{mL\cosh\theta},\\
&\kappa(\theta)=\int_{-\infty}^{\infty}\frac{d\theta_0}{2\pi}\frac{1}{\cosh(\theta-\theta_0)}\log\left(\frac{1-e^{-mL\cosh\theta_0}}{1+e^{-mL\cosh\theta_0}}\right).
\ea\ee
The energy of an $N$-particle state with finite size $L$ is given by 
\be
E_{NS}^N(L) = E_{NS}^0(L)+\sum_{i=1}^N\omega_{k_i}(L),\quad
E_{R}^N(L) = E_{R}^0(L)+\sum_{i=1}^N\omega_{k_i}(L),
\ee
with fermion energy $\omega_{k_i}(R)=\sqrt{m^2+(2\pi k/R)^2}$. The ground state energies for each sector ($|0\rangle_{NS}$ and $|0\rangle_R$) are
\be\ba
E_{NS}^0(L) = e(m,\lambda=0)L-m\int_{-\infty}^{\infty}\frac{d\theta}{2\pi}\cosh\theta\log(1+e^{-mL\cosh\theta}),\\
E_{R}^0(L) = e(m,\lambda=0)L-m\int_{-\infty}^{\infty}\frac{d\theta}{2\pi}\cosh\theta\log(1-e^{-mL\cosh\theta}),\\
\ea\ee
with the energy density $e(m,\lambda=0)=(m^2/8\pi)\log m^2$.

\section{Wavefunction and DSF near the integrable point}
\label{app:FFPT}
In this appendix, we calculate the DSF near the integrable point.
The infinite volume resolution of identity for IIFT can be written as
\be\ba
\mathbb{I} = |0\rangle\langle0|+\sum_{N=1}^{\infty}\sum_{\{a_N\}}\frac{1}{\prod_{a=1}^sk_a^{(N)}!}\left(\prod_{i=1}^N\int_{-\infty}^{\infty}\frac{d\theta_i}{2\pi}\right)|\theta_1\cdots\theta_N\rangle\langle\theta_1\cdots\theta_N|,
\label{eq:id}
\ea\ee
where $\{a_N\}$ denotes the configuration of $N$-particle states and $k_a^{N}$ counts the number of $a$-type particles in the set.
It is useful to consider the theory in a large but finite volume, 
where the rapidities are discretized through the Bethe-Yang equation 
\be
\mc{Q}_i = mL\sinh\theta_i+\sum_{j\neq i}(-i)\log S(\theta_i-\theta_j)=2\pi I_i,
\label{eq:discrete}
\ee
with integer $I_i$ and scattering matrix $S(\theta)$.
The integration in Eq.~\eqref{eq:id} is discretized as 
\be\ba
&\prod_{i=1}^N\int_{-\infty}^{\infty}\frac{d\theta_i}{2\pi}=\prod_{i=1}^N\left.\sum_{\{\theta_i^a\}}\frac{1}{\rho_{a_1,\cdots a_N}(\theta_1,\cdots,\theta_N)}\right|_{\{\theta\}=\{\theta^a\}},\\
&\rho_{a_1,\cdots a_N}(\theta_1,\cdots,\theta_N) = \det\left(\frac{d\mc{Q}_k}{d\theta_l}\right),\quad k,l = 1,\cdots,N,
\ea\ee
with rapidities $\{\theta^a\}$ solved from Eq.~\eqref{eq:discrete}.
Introduce the finite volume notation 
\be
|\theta_1\cdots\theta_N\rangle_L=\frac{|\theta_1\cdots\theta_N\rangle}{\sqrt{\rho_{a_1,\cdots a_N}(\theta_1,\cdots,\theta_N)}}
\ee
with normalization condition
\be 
_L\langle\theta_1\cdots\theta_L|\theta_1\cdots\theta_L\rangle_L\frac{1}{\prod_{a=1}^sk_a^{(N)}!}=1.
\ee

First order perturbation of the ground state wave function is obtained as 
\be\ba
&|\tilde{0}\rangle \simeq |0\rangle+\sum_{N=1}^{\infty}\sum_{\{a_N\}}\frac{1}{\prod_{a=1}^sk_a^{(N)}!}\left(\prod_{i=1}^N\int_{-\infty}^{\infty}\frac{d\theta_i}{2\pi}\right)\langle \theta_1\cdots\theta_N|\rho\int dx\cos\phi(x)|0\rangle|\theta_1\cdots\theta_N\rangle\frac{-1}{E_N}\\
&=|0\rangle+\sum_{N=1}^{\infty}\sum_{\{a_N\}}\left(\prod_{i=1}^N\int_{-\infty}^{\infty}\frac{d\theta_i}{2\pi}\right)\delta\left(\sum_{i=1}^Nm_i^0\sinh\theta_i\right)\rho\langle\theta_1\cdots\theta_N|\cos\phi(0)|0\rangle|\theta_1\cdots\theta_N\rangle\frac{-1}{E_N}\\
&=|0\rangle-\rho\left\{\sum_{i=2,4,6}\frac{\langle B_i(0)|\cos\phi|0\rangle}{(m_{B_i}^0)^2}|B_i(0)\rangle+\int\frac{d\theta}{2\pi}\frac{\langle B_1(\theta)B_1(-\theta)|\cos\phi|0\rangle}{2(m_{B_1}^0\cosh\theta)^2}|B_1(\theta)B_1(-\theta)\rangle\right.\\
&+\left.\int\frac{d\theta}{2\pi}\frac{\langle A_{+1}(\theta)A_{-1}(-\theta)|\cos\phi|0\rangle}{2(m_{A_{\pm1}}^0\cosh\theta)^2}|A_{+1}(\theta)A_{-1}(-\theta)\rangle+\cdots\cdots\right\}
\label{eq:gs}
\ea\ee
where $\cdots\cdots$ denotes other multiparticle states that are not $\mc{C}$-odd.

And for $B_1$,
\be\ba
&|\tilde{B}_1(0)\rangle\simeq|B_1(0)\rangle-\rho\left\{\sum_{i=3,5}\frac{\langle B_i(0)|\cos\phi|B_1(0)\rangle}{m_{B_i}^0(m_{B_i}^0-m_{B_1}^0)}|B_i(0)\rangle\right.\\
&\left.+\int_{-\infty}^{\infty}\frac{d\theta}{2\pi}\frac{\langle B_{1}(\theta)B_{2}(\theta^{\prime})|\cos\phi|B_1(0)\rangle}{(m_{B_2}^0\cosh\theta^{\prime})(m_{B_1}^0\cosh\theta+m_{B_2}^0\cosh\theta^{\prime}-m_{B_1}^0)}|B_{1}(\theta)B_{2}(\theta^{\prime})\rangle+\cdots\right\}
\label{eq:B1}
\ea\ee
with $\theta^{\prime}=\text{arcsinh}(m_{B_1}^0/(m_{B_2}^0\sinh\theta))$ and other $\mc{C}$-odd components not listed.

Similar to the selection rules for the ground state,
for $B_2$ we have 
\be\ba
&|\tilde{B}_2(0)\rangle\simeq|B_2(0)\rangle-\rho\left\{\frac{\langle 0|\cos\phi|B_2(0)\rangle}{-m_{B_2}^0}\delta(0)|0\rangle+\sum_{i=4,6}\frac{\langle B_i(0)|\cos\phi|B_2(0)\rangle}{m_{B_i}^0(m_{B_i}^0-m_{B_2}^0)}|B_i(0)\rangle\right.\\
&\left.+\int_{-\infty}^{\infty}\frac{d\theta}{2\pi}\frac{\langle B_1(\theta)B_{1}(-
\theta)|\cos\phi|B_2(0)\rangle}{m_{B_1}^0\cosh\theta(2m_{B_1}^0\cosh\theta-m_{B_2}^0)}|B_{1}(\theta)B_{1}
(\theta)\rangle+\cdots\right\}.
\label{eq:B2}
\ea\ee
The $\delta(0)$ ``divergence'' will be cancelled in the calculation of the DSF.
To verify this explicitly,
we rewrite Eq.~\eqref{eq:B2} in the finite size case 
\be\ba 
&\sqrt{\tilde{m}_{B_2}L}|\tilde{B}_2(0)\rangle_L = \sqrt{m_{B_2}^0L}\left\{|B_2(0)\rangle_L-\rho L\left[\frac{_L\langle 0|\cos\phi|B_2(0)\rangle_L}{-m_{B_2}^0}|0\rangle_L\right.\right.\\
&\left.\left.+\sum_{i=4,6}\frac{_L\langle B_i(0)|\cos\phi|B_2(0)\rangle_L}{m_{B_i}^0-m_{B_2}^0}|B_i(0)\rangle_L
+\sum_{\{\theta^a\}}\frac{_L\langle B_{1}(\theta^a)B_{1}(-\theta^a)|\cos\phi|B_2(0)\rangle_L}{m_{B_1}^0L\cosh\theta^a (2m_{B_{1}}\cosh\theta^a-m_{B_2}^0)}\right.\right.\\
&\times\left.\left.|B_{1}(\theta^a)B_{1}(-\theta^a)\rangle_L+\cdots\right]\right\}.\\
\ea\ee
And for $A_{\pm1}$,
\be\ba
|\tilde{A}_{\pm}(0)\rangle\simeq|A_{\pm}(0)\rangle-\rho\int_{-\infty}^{\infty}\frac{d\theta}{2\pi}\frac{\langle A_{\pm 1}(\theta)B_1(\theta^{\prime})|\cos\phi|A_{\pm1}(0)\rangle}{m_{A_{\pm1}}\cosh\theta^{\prime}(m_{A_{\pm 1}}^0\cosh\theta+m_{B_1}^0\cosh\theta^{\prime}-m_{A_{\pm1}^0})}|A_{\pm1}(\theta)B_1(-\theta)\rangle+\cdots.
\ea\ee
with $\theta^{\prime}=\text{arcsinh}(m_{B_1}^0/(m_{A_{\pm1}}^0\sinh\theta))$.

The single particle contribution in the zero-momentum DSF of $\epsilon$ is obtained as 
\be\ba
D_{\text{1p}}^{\epsilon}(\omega,q=0)&=\iint_{-\infty}^{\infty}dx dt\int_{-\infty}^{\infty}\frac{d\theta}{2\pi} |\langle \tilde{0}|\cos\phi+\cos\tilde{\phi}|\tilde{P_r}(\theta)\rangle |^2e^{it(\tilde{E}_0-\tilde{E}_{r}(\theta))}e^{i\tilde{Q}_r(\theta)x}e^{i\omega t}e^{-iqx}\\
&=\int_{-\infty}^{\infty}dte^{i\omega t}\int_{-\infty}^{\infty}\frac{d\theta}{2\pi}e^{it(\tilde{E}_0-\tilde{E}_r(\theta))}\delta(\tilde{Q}_r(\theta))|\langle \tilde{0}|\cos\phi+\cos\tilde{\phi}|\tilde{P}_r(\theta)\rangle|^2\\
&=\delta(\omega+\tilde{E}_0-\tilde{E}_r(0))\left(\left.\frac{d\tilde{Q}_r}{d\theta}\right|_{\theta=0}\right)^{-1}|\langle \tilde{0}|\cos\phi+\cos\tilde{\phi}|\tilde{P}_r(0)\rangle|^2
\label{eq:d1p}
\ea\ee
where the perturbed single particles have been inserted.
$\tilde{E}_r$, $\tilde{Q}_r$ denotes the perturbed energy and momentum for particle $\tilde{P}_r$, where we keep to the first order in $\rho$.
$(\tilde{E}_r(0)-\tilde{E}_0)^2=m_r^2+\delta m_r^2$ is obtained in the main text, and we have
\be\ba
&\tilde{Q}_r(\theta)=\langle \tilde{P}_r(\theta)|\hat{P}|\tilde{P}_r(\theta)\rangle\\
&=m_{r}^0\cosh\theta+\rho^2\sum_{\{P_j\}}\langle P_1(\theta_1)\cdots P_m(\theta_m)|\hat{P}|P_1(\theta_1)\cdots P_m(\theta_m)\rangle+\mc{O}(\lambda^4)
\ea\ee
with $\{P_j\}$ the set of (multi-)particle components in $|\tilde{P}_r\rangle$.
In this way, 
$d\tilde{Q}_r/d\theta=m_r^0\sinh\theta$.

Consider the transition element in Eq.~\eqref{eq:d1p},
it can be directly obtained by inserting previously obtained wave functions and calculating each term using the form factors obtained from IIFT.
For the $\delta(0)$ term in Eq.~\eqref{eq:B2},
it will be cancelled by a disconnected part when calculating the transition element.
The order of divergent is more transparent by considering the finite size formula, \emph{i.e.}, 
\be\ba
&\langle \tilde{0}|\cos\phi+\cos\tilde{\phi}|\tilde{B}_2(0)\rangle=_L\langle \tilde{0}|\cos\phi+\cos\tilde{\phi}|\tilde{B}_2(0)\rangle_L\sqrt{\tilde{m}_{B_2}L}\\
&=\sqrt{m^0_{B_2}L}\left\{ _L\langle0|\cos\phi|B_2(0)\rangle_L-\rho L\frac{_L\langle0|\cos\phi|B_2(0)\rangle_L}{-m_{B_2}^0}\,_L\langle 0|\cos\phi|0\rangle_L\right.\\
&-\rho L \sum_{i=2,4,6}\frac{_L\langle B_i(0)|\cos\phi|0\rangle_L}{m_{B_i}^0}\,_L\langle B_2(0)|\cos\phi|B_i(0)\rangle_L\left.+\cdots\right\}
\label{eq:ele}
\ea\ee
where the $i=2$ case in the third term contains a disconnected piece, \emph{i.e.}, 
\be 
_L\langle B_2(0)|\cos\phi|B_2(0)\rangle_L=\frac{1}{m_{B_2}L}\langle 0|\cos\phi|B_2(i\pi)B_2(0)\rangle+\langle 0|\cos\phi|0\rangle.
\label{eq:disconnect}
\ee
The second term in Eq.~\eqref{eq:disconnect} cancels the second term in Eq.~\eqref{eq:ele}. As a result,
\be\ba
&\langle \tilde{0}|\cos\phi+\cos\tilde{\phi}|\tilde{B}_2(0)\rangle=\langle0|\cos\phi|B_2(0)\rangle-\rho\sum_{i=4,6}\frac{\langle B_i(0)|\cos\phi|0\rangle}{(m_{B_i}^0)^2}\langle B_2(0)|\cos\phi|B_i(0)\rangle\\
&-\rho\sum_{i=4,6}\frac{\langle B_i(0)|\cos\phi|B_2(0)\rangle}{m_{B_i}^0(m_{B_i}^0-m_{B_2}^0)}\langle 0|\cos\phi|B_i(0)\rangle\\
&-\rho \int_{-\infty}^{\infty}\frac{d\theta}{2\pi}\frac{\langle B_1(\theta)B_1(-\theta)|\cos\phi|B_2(0)\rangle}{m_{B_1}^0\cosh\theta(2m_{B_1}^0\cosh\theta-m_{B_2}^0)}\langle 0|\cos\phi|B_1(\theta)B_1(-\theta)\rangle+\cdots.
\ea\ee

\section{Truncated conformal space approach}
\label{app:TCSA}
The integrable model starts from two weakly coupled quantum critical TFICs.
Each of the quantum critical TFIC can be described by a $c=1/2$ CFT.
In this section we generalize the truncated conformal space method introduced 
in \cite{yurov1990,yurov1991} to odd fermion number parity
for obtaining full mass spectrum of the Ising$_h^2$ particles and their classification.
The basic set up is summarized as below.

The conformal-invariant Hamiltonian with central charge $c=1/2$ can be organized as 
\be\ba
&H_{\text{c=1/2}}=\frac{2\pi}{L}\sum_{n=0}^{\infty}\left(n+\frac{1}{2}\right)[a_{-n-1/2}a_{n+1/2}+\overline{a}_{-n-1/2}\overline{a}_{n+1/2}]-\frac{1}{24}\quad (\text{NS-sector}),\\
&H_{\text{c=1/2}}=\frac{2\pi}{L}\sum_{n=0}^{\infty}n(a_{-n} a_{n}+\overline{a}_{-n}\overline{a}_{n})+\frac{1}{12}\quad (\text{R-sector}),
\label{eq:cft}
\ea\ee
with $a_n$ and $\overline{a}_n$ denoting the fermion operators from holomorphic and anti-holomorphic parts.
Following anticommutation relations are satified: $\{a_{n+1/2},a_{m-1/2}\}=\delta_{m+n}$, $\{\overline{a}_{n+1/2},\overline{a}_{m-1/2}\}=\delta_{m+n}$; $\{a_m,a_n\}=\delta_{m+n}$, $\{\overline{a}_{n},\overline{a}_{m}\}=\delta_{m+n}$ and $\{a_n,\overline{a}_m\}=0$.
Correspondingly,
the momentum operator for the two sectors is 
\be
\hat{Q}=\sum_{n=0}^{\infty}\left(n+\frac{1}{2}\right)[a_{-n-1/2}a_{n+1/2}-\overline{a}_{-n-1/2}\overline{a}_{n+1/2}]
\ee
with $n$ being integer and half-integer for the NS and R sectors, respectively.

Following \cite{yurov1991},
the NS-sector ($\mc{F}_+$) owns a unique vacuum $|I\rangle$ and can be 
further divided into two subsectors,
depending on whether the total number of fermion creation operators is odd ($\mc{F}_{\varepsilon}$)  or even ($\mc{F}_{I}$).
And the R-sector $\mc{F}_{-}=\mc{F}_{\sigma}\oplus\mc{F}_{\mu}$,
with $\mc{F}_{\sigma}$ containing even number of fermion creation operators from both holomorphic
 and anti-holomorphic parts, and $\mc{F}_{\mu}$ containing odd of them from both parts.
$\mc{F}_{\sigma}$ and $\mc{F}_{\mu}$ are built from their corresponding vacuum $|\sigma\rangle$ and $|\mu\rangle$, respectively.
$\mc{F}_{\varepsilon}$ and $\mc{F}_{\mu}$ are connected with the odd fermion number sector in the massive case,
which are disconnected with $\mc{F}_{I}$ and $\mc{F}_\sigma$ sectors.

A set of fermions ($c_n$ and $\overline{c}_n$) is introduced to calculate the matrix entries, 
which connect the NS sector in the infinite past and the R-sector in the infinite future.
They are related to the normal fermions as
\be\ba
&a_{n-1/2}=\sum_{k=0}^{\infty}z_0^{-k-1/2}Q_kc_{n+k},\quad 
\overline{a}_{n-1/2}=\sum_{k=0}^{\infty}\overline{z}_0^{k+1/2}Q_k\overline{c}_{n+k},\\
&a_n=i\sum_{k=0}^{\infty}z_0^kQ_kc_{n-k},\quad 
\overline{a}_n=-i\sum_{k=0}^{\infty}\overline{z}_0^kQ_k\overline{c}_{n-k},
\label{eq:relation}
\ea\ee
with $Q_k=(2k-1)!!/(2^kk!)$ and $z_0$ related to the position of $\sigma$ which is taken as $1$.
In this way, the $c$ fermions commute with $\sigma(0,0)$ \cite{yurov1991}.
The anticommutation relations read
\be
\{c_m,c_n\}=z_0\delta_{m+n-1}-\delta_{m+n},\quad 
\{\overline{c}_m,\overline{c}_n\}=\overline{z}_0\delta_{m+n-1}-\delta_{m+n}.
\ee
For the 
Ising ladder Hamiltonian $H_{\text{II}}(0,0)$ as a perturbation for the coupled CFTs (labeled as $H_{\text{IIFT}}^f$),
we have the following combinations of the Hilbert space $\mc{F}_+^{(1)}\otimes\mc{F}_+^{(2)}$,
$\mc{F}_-^{(1)}\otimes\mc{F}_-^{(2)}$, $\mc{F}_+^{(1)}\otimes\mc{F}_-^{(2)}$ and $\mc{F}_-^{(1)}\otimes\mc{F}_+^{(2)}$.
The last term in Eq.~\eqref{eq:action} acts between the first two sectors or the last two sectors, leaving them independent from each other.
For simplicity, here we consider the first two subspaces with zero total momentum (Tab.~\ref{tab:basis}).
Elements of diagonal part in $H_{\text{IIFT}}^f$ are given by directly counting the scaling dimensions 
(referred to as "level") as shown in Eq.~\eqref{eq:cft}.
It remains to calculate 
\be\ba
\langle F_-^{i}|\sigma^{(1)}(0,0)|F_+^{\alpha}\rangle\langle F_-^{j}|\sigma^{(2)}(0,0)|F_+^{\beta}\rangle,
\label{eq:interpolation}
\ea\ee
with  both $F_+^{\alpha},F_+^{\beta} \in \mc{F}_I$ or $\mc{F}_\varepsilon$, and both $F_-^{i}, F_-^j \in \mc{F}_{\sigma}$ or $\mc{F}_{\mu}$.
The transition elements are calculated by arranging the fermion operator in the basis 
in a lowering order of scaling dimensions from the right.
Notice that
\be\ba 
a_{k-1/2}|I\rangle=\overline{a}_{k-1/2}|I\rangle=0\quad\text{for}\, k>0,\\
\langle0|a_k=\,\langle0|\overline{a}_k=0,\quad\text{for}\, k<0,
\ea\ee
with $|0\rangle$ denoting the R-sector ground state without further distinguishing the degeneracy.
Inserting Eq.~\eqref{eq:relation}, these implies 
\be\ba
c_k|I\rangle=\overline{c}_k|I\rangle=0\quad\text{for}\, k>0,\\
\langle0|c_k=\,\langle0|\overline{c}_k=0,\quad\text{for}\, k<0.
\label{eq:constrain}
\ea\ee

\begin{table}[h]
    \caption{Examples of basis for the first 5 levels with zero total momentum. $(a;b)$ denotes 
    the state with operator $a$ and $b$ operating on the corresponding vacua for the two TFICs. $/$ stands for no state.}
      \vspace{3pt}
    \centering 
    \begin{tabular}{ccccc}
        \toprule 
        level  & $\mc{F}_I\mc{F}_I$ & $\mc{F}_\varepsilon\mc{F}_\varepsilon$ & $\mc{F}_\sigma\mc{F}_\sigma$ & $\mc{F}_\mu\mc{F}_\mu$\\
        \toprule
        0  &  & / & &$(\overline{a}_0a_0;\overline{a}_0a_0)$\\
        \midrule
        1  & $(\overline{a}_{-\frac{1}{2}}a_{-\frac{1}{2}};)$ & $(a_{-\frac{1}{2}};\overline{a}_{\frac{1}{2}})$ & /& / \\
        \midrule
        2  & $(\overline{a}_{-\frac{1}{2}}a_{-\frac{1}{2}};\overline{a}_{-\frac{1}{2}}a_{-\frac{1}{2}})$ & / & $(\overline{a}_{-1}\overline{a}_0a_{-1}a_0;)$& \makecell{$(\overline{a}_0a_0;\overline{a}_{-1}a_{-1})$,\\ $(\overline{a}_0a_{-1};\overline{a}_{-1}a_0)$}\\
        \midrule
        3  & $(\overline{a}_{-\frac{3}{2}}a_{-\frac{3}{2}};)$ & $(a_{-\frac{3}{2}};\overline{a}_{\frac{3}{2}})$ & /& /\\
        \midrule
        4 & \makecell{$(\overline{a}_{-\frac{3}{2}}\overline{a}_{-\frac{1}{2}}a_{-\frac{3}{2}}a_{-\frac{1}{2}};)$,\\ ($a_{-\frac{3}{2}}a_{-\frac{1}{2}};\overline{a}_{-\frac{3}{2}}\overline{a}_{-\frac{1}{2}}$),\\ $(\overline{a}_{-\frac{3}{2}}a_{-\frac{3}{2}};\overline{a}_{-\frac{1}{2}}a_{-\frac{1}{2}})$,\\ $(\overline{a}_{-\frac{3}{2}}a_{-\frac{1}{2}};\overline{a}_{-\frac{1}{2}}a_{-\frac{3}{2}})$}& \makecell{$(a_{-\frac{3}{2}};\overline{a}_{-\frac{3}{2}}\overline{a}_{-\frac{1}{2}}a_{-\frac{1}{2}})$\\$(a_{-\frac{1}{2}};\overline{a}_{-\frac{3}{2}}\overline{a}_{-\frac{1}{2}}a_{-\frac{3}{2}})$,\\$(\overline{a}_{-\frac{3}{2}};\overline{a}_{-\frac{1}{2}}a_{-\frac{3}{2}}a_{-\frac{1}{2}})$,\\$(\overline{a}_{-\frac{3}{2}}a_{-\frac{3}{2}}a_{-\frac{1}{2}};\overline{a}_{-\frac{1}{2}})$}&\makecell{$(\overline{a}_{-2}\overline{a}_0a_{-2}a_0;)$,\\ $(a_{-2}a_0;\overline{a}_{-2}\overline{a}_0)$,\\ $(a_{-1}a_0;\overline{a}_{-2}\overline{a}_0a_{-1}a_0)$,\\ $(\overline{a}_{-1}\overline{a}_0;\overline{a}_{-1}\overline{a}_0a_{-2}a_0)$,\\ $(\overline{a}_{-1}\overline{a}_0a_{-1}a_0;\overline{a}_{-1}\overline{a}_0a_{-1}a_0)$ }&
        \makecell{$(\overline{a}_0a_0;\overline{a}_{-2}a_{-2})$,\\ $(\overline{a}_0a_{-2};\overline{a}_{-2}a_{0})$,\\$(\overline{a}_0a_{-1};\overline{a}_{-2}a_{-1})$,\\ $(\overline{a}_{-1}a_0;\overline{a}_{-1}a_{-2})$,\\ $(\overline{a}_{-1}a_{-1};\overline{a}_{-1}a_{-1})$}\\
        \bottomrule
    \end{tabular}
    \label{tab:basis}
\end{table}

First we expand $a$ fermions in terms of $c$ fermions.
This can be done with finite numebr of $c$ fermions by putting the constraint in Eq.~\eqref{eq:constrain} from both the bra and ket vacua.
Also notice that the $c$ and $\overline{c}$ parts are factorized in the calculation. 
If $F_+^{\alpha}\in\mc{F}_{I}$ and $F_-^{i}\in\mc{F}_{\sigma}$,
$\langle F_-^{i}|\sigma^{(1)}(0,0)|F_+^{\alpha}\rangle$ can be reduced to a multiplier of either $\langle0|\sigma(0,0)|I\rangle$ or $\langle0|c_0\overline{c}_0\sigma(0,0)|I\rangle$.
If $F_+^{\alpha}\in\mc{F}_{\varepsilon}$ and $F_-^{i}\in\mc{F}_{\mu}$,
it can be reduced to a multiplier of either $\langle0|c_0\sigma(0,0)|I\rangle$ or $\langle0|\overline{c}_0\sigma(0,0)|I\rangle$.
Here we take the convention that $\langle0|c_0\overline{c}_0\sigma(0,0)|I\rangle=\frac{1}{2}\langle0|\sigma(0,0)|I\rangle$, 
$\langle0|c_0\sigma(0,0)|I\rangle=\frac{i}{\sqrt{2}}\langle0|\sigma(0,0)|I\rangle$,
$\langle0|\overline{c}_0\sigma(0,0)|I\rangle=\frac{-i}{\sqrt{2}}\langle0|\sigma(0,0)|I\rangle$.
Actually, the sign and $i$ here do not affect the spectrum as will be clear later.
The overall normalization for these elements in the CFT follows 
\be
\langle0|\sigma(0,0)|I\rangle=\left(\frac{2\pi}{L}\right)^{1/8}.
\ee
Then the block matrix of the Hamiltonian follows by
\be
H_{\text{IIFT}}^{f,e}=\left(
    \begin{matrix}
    H^{(1)}_{\text{c=1/2}}\otimes H^{(2)}_{\text{c=1/2}}(\mc{F}_I,\mc{F}_I)&\lambda L\left[\langle F_+^{\alpha},F_+^{\beta}|\sigma^{(1)}\sigma^{(2)}|F_-^i,F_-^j\rangle\right]\\
    \lambda L\left[\langle F_-^i,F_-^j|\sigma^{(1)}\sigma^{(2)}| F_+^{\alpha},F_+^{\beta}\rangle\right] & H^{(1)}_{\text{c=1/2}}\otimes H^{(2)}_{\text{c=1/2}}(\mc{F}_\sigma,\mc{F}_\sigma)
    \end{matrix}
\right),
\ee
with $F_+^{(\cdot)}\in\mc{F}_I$, $F_-^{(\cdot)}\in\mc{F}_\sigma$ and $[\,]$ denoting the matrix,
and
\be
H_{\text{IIFT}}^{f,o}=\left(
    \begin{matrix}
    H^{(1)}_{\text{c=1/2}}\otimes H^{(2)}_{\text{c=1/2}}(\mc{F}_\varepsilon,\mc{F}_\varepsilon)&\lambda L\left[\langle F_+^{\alpha},F_+^{\beta}|\sigma^{(1)}\sigma^{(2)}|F_-^i,F_-^j\rangle\right]\\
    \lambda L\left[\langle F_-^i,F_-^j|\sigma^{(1)}\sigma^{(2)}| F_+^{\alpha},F_+^{\beta}\rangle\right] & H^{(1)}_{\text{c=1/2}}\otimes H^{(2)}_{\text{c=1/2}}(\mc{F}_\mu,\mc{F}_\mu)
    \end{matrix}
\right),
\label{eq:oddmat}
\ee 
with $F_+^{(\cdot)}\in\mc{F}_\varepsilon$ and $F_-^{(\cdot)}\in\mc{F}_\mu$.
Furthermore,
$F_-^i$  in Eq.~\eqref{eq:oddmat} with an odd number of $\overline{a}_n$ and an even number of $a_n$ will lead to matrix element containing $\langle0|\overline{c}_0\sigma(0,0)|I\rangle$ in the off-diagonal block, 
and vice versa.
Explicitly,
Eq.~\eqref{eq:oddmat} is organized as following
\be
\left(\begin{array}{c|c}
\begin{matrix}
    e_1 &0 &0 & 0& \cdots\\
    0& e_2 &0 &0 &\cdots\\
    0& 0& e_3 &0 &\cdots\\
    0& 0& 0 & e_4 &\cdots\\
    0& 0& 0 & 0 &\ddots\\
\end{matrix} &
\begin{array}{c|c|c|c}
    & & &  \\
    & &\cdots &  \\
    & & &  \\
    & & &  \\
    & & &  \\
\end{array}\\ \hline
\begin{matrix}
& & & &\\ \hline
& & &\vdots &\\ \hline
& & & &\\ \hline
\end{matrix} &
\begin{matrix}
e_1^{\prime} & 0 & \cdots\\
0 & e_2^{\prime} & \cdots \\
0&0 & \ddots
\end{matrix}
\end{array}\right).
\ee
Changing sign in both $\langle0|\overline{c}_0\sigma(0,0)|I\rangle$ and $\langle0|c_0\sigma(0,0)|I\rangle$ does not modify the matrix elements.
However, changing the sign of $\langle0|\overline{c}_0\sigma(0,0)|I\rangle$ or $\langle0|c_0\sigma(0,0)|I\rangle$
will result in additional minus sign in some columns of right-upper block and the corresponding rows in the left-lower block.
These lead to additional minus sign in the corresponding positions of the original eigenvector,
with the eigenvalue unchanged.
The sign-change operation can also be generalized to a phase factor.

Moreover, 
$H_{\text{IIFT}}^{f,e}$ and $H_{\text{IIFT}}^{f,o}$ can be further block-diagonalized in the subspace that is odd/even w.r.t. exchanging the two chains,
as in the massive case.

\bibliographystyle{JHEP}
\bibliography{Refs}

\providecommand{\href}[2]{#2}\begingroup\raggedright\begin{thebibliography}{10}

\bibitem{PhysRevD.10.2445}
K.G.~Wilson, \emph{Confinement of quarks},
  \href{https://doi.org/10.1103/PhysRevD.10.2445}{\emph{Phys. Rev. D}
  {\bfseries 10} (1974) 2445}.

\bibitem{tHooft:1974pnl}
G.~'t~Hooft, \emph{{A Two-dimensional model for mesons}},
  \href{https://doi.org/10.1016/0550-3213(74)90088-1}{\emph{Nucl. Phys. B}
  {\bfseries 75} (1974) 461}.

\bibitem{Abdalla:1995dm}
E.~Abdalla and M.C.B.~Abdalla, \emph{{Updating QCD in two-dimensions}},
  \href{https://doi.org/10.1016/0370-1573(95)00019-4}{\emph{Phys. Rept.}
  {\bfseries 265} (1996) 253}
  [\href{https://arxiv.org/abs/hep-th/9503002}{{\ttfamily hep-th/9503002}}].

\bibitem{Fateev:2009jf}
V.A.~Fateev, S.L.~Lukyanov and A.B.~Zamolodchikov, \emph{{On mass spectrum in
  't Hooft's 2D model of mesons}},
  \href{https://doi.org/10.1088/1751-8113/42/30/304012}{\emph{J. Phys. A}
  {\bfseries 42} (2009) 304012}
  [\href{https://arxiv.org/abs/arXiv:0905.2280}{{\ttfamily arXiv:0905.2280}}].

\bibitem{Ambrosino:2023dik}
F.~Ambrosino and S.~Komatsu, \emph{{2d QCD and Integrability, Part I: 't Hooft
  model}},  \href{https://arxiv.org/abs/arXiv:2312.15598}{{\ttfamily
  arXiv:2312.15598}}.

\bibitem{Ambrosino:2024prz}
F.~Ambrosino and S.~Komatsu, \emph{{2d QCD and Integrability, Part II:
  Generalized QCD}},  \href{https://arxiv.org/abs/arXiv:2406.11078}{{\ttfamily
  arXiv:2406.11078}}.

\bibitem{Litvinov:2024riz}
A.~Litvinov and P.~Meshcheriakov, \emph{{Meson mass spectrum in QCD2 't Hooft's
  model}}, \href{https://doi.org/10.1016/j.nuclphysb.2024.116766}{\emph{Nucl.
  Phys. B} {\bfseries 1010} (2025) 116766}
  [\href{https://arxiv.org/abs/arXiv:2409.11324}{{\ttfamily
  arXiv:2409.11324}}].

\bibitem{PhysRevD.18.1259}
B.M.~McCoy and T.T.~Wu, \emph{Two-dimensional {Ising} field theory in a
  magnetic field: Breakup of the cut in the two-point function},
  \href{https://doi.org/10.1103/PhysRevD.18.1259}{\emph{Phys. Rev. D}
  {\bfseries 18} (1978) 1259}.

\bibitem{Fonseca2003}
P.~Fonseca and A.~Zamolodchikov, \emph{{Ising} field theory in a magnetic
  field: {Analytic} properties of the free energy},
  \href{https://doi.org/10.1023/A:1022147532606}{\emph{J. Stat. Phys.}
  {\bfseries 110} (2003) 527}
  [\href{https://arxiv.org/abs/hep-th/0112167}{{\ttfamily hep-th/0112167}}].

\bibitem{zam2006}
P.~Fonseca and A.~Zamolodchikov, \emph{Ising {Spectroscopy} {I}: {Mesons} at
  ${T < T_c}$},  \href{https://arxiv.org/abs/hep-th/0612304}{{\ttfamily
  hep-th/0612304}}.

\bibitem{Zamolodchikov:2013ama}
A.~Zamolodchikov, \emph{{Ising Spectroscopy II: Particles and poles at $T
  >T_c$}},  \href{https://arxiv.org/abs/arXiv:1310.4821}{{\ttfamily
  arXiv:1310.4821}}.

\bibitem{Xu:2022mmw}
H.L.~Xu and A.~Zamolodchikov, \emph{{2D {Ising} Field Theory in a magnetic
  field: the Yang-Lee singularity}},
  \href{https://doi.org/10.1007/JHEP08(2022)057}{\emph{J. High Energy Phys.}
  {\bfseries 08} (2022) 057}
  [\href{https://arxiv.org/abs/arXiv:2203.11262}{{\ttfamily
  arXiv:2203.11262}}].

\bibitem{Xu:2023nke}
H.L.~Xu and A.~Zamolodchikov, \emph{{Ising field theory in a magnetic field:
  \ensuremath{\varphi}$^{3}$ coupling at T \ensuremath{>} T$_{c}$}},
  \href{https://doi.org/10.1007/JHEP08(2023)161}{\emph{J. High Energy Phys.}
  {\bfseries 08} (2023) 161}
  [\href{https://arxiv.org/abs/arXiv:2304.07886}{{\ttfamily
  arXiv:2304.07886}}].

\bibitem{Mussardo:2010sy}
G.~Mussardo, \emph{{Integrability, non-integrability and confinement}},
  \href{https://doi.org/10.1088/1742-5468/2011/01/P01002}{\emph{J. Stat. Mech.}
  {\bfseries 1101} (2011) P01002}
  [\href{https://arxiv.org/abs/arXiv:1010.5519}{{\ttfamily arXiv:1010.5519}}].

\bibitem{DELFINO2008265}
G.~Delfino and P.~Grinza, \emph{Confinement in the q-state potts field theory},
  \href{https://doi.org/https://doi.org/10.1016/j.nuclphysb.2007.09.003}{\emph{Nucl.
  Phys. B} {\bfseries 791} (2008) 265}
  [\href{https://arxiv.org/abs/arXiv:0706.1020}{{\ttfamily arXiv:0706.1020}}].

\bibitem{Lencses:2021igo}
M.~Lencs\'es, G.~Mussardo and G.~Tak\'acs, \emph{{Confinement in the
  tricritical {Ising} model}},
  \href{https://doi.org/10.1016/j.physletb.2022.137008}{\emph{Phys. Lett. B}
  {\bfseries 828} (2022) 137008}
  [\href{https://arxiv.org/abs/arXiv:2111.05360}{{\ttfamily
  arXiv:2111.05360}}].

\bibitem{DELFINO1998}
G.~Delfino and G.~Mussardo, \emph{Non-integrable aspects of the multi-frequency
  sine-{Gordon} model},
  \href{https://doi.org/https://doi.org/10.1016/S0550-3213(98)00063-7}{\emph{Nucl.
  Phys. B} {\bfseries 516} (1998) 675}
  [\href{https://arxiv.org/abs/hep-th/9709028}{{\ttfamily hep-th/9709028}}].

\bibitem{Roy:2023byz}
A.~Roy and S.L.~Lukyanov, \emph{{Soliton confinement in a quantum circuit}},
  \href{https://doi.org/10.1038/s41467-023-43107-3}{\emph{Nat. Commun.}
  {\bfseries 14} (2023) 7433}
  [\href{https://arxiv.org/abs/arXiv:2302.06289}{{\ttfamily
  arXiv:2302.06289}}].

\bibitem{Rutkevich:2023lzj}
S.~Rutkevich, \emph{{Soliton confinement in the double sine-Gordon model}},
  \href{https://doi.org/10.21468/SciPostPhys.16.2.042}{\emph{SciPost Phys.}
  {\bfseries 16} (2024) 042}
  [\href{https://arxiv.org/abs/arXiv:2311.07303}{{\ttfamily
  arXiv:2311.07303}}].

\bibitem{PhysRevD.23.1862}
M.~Stone, \emph{Bound states in {Ising} field theory},
  \href{https://doi.org/10.1103/PhysRevD.23.1862}{\emph{Phys. Rev. D}
  {\bfseries 23} (1981) 1862}.

\bibitem{PhysRevLett.77.2790}
H.J.~Schulz, \emph{Dynamics of coupled quantum spin chains},
  \href{https://doi.org/10.1103/PhysRevLett.77.2790}{\emph{Phys. Rev. Lett.}
  {\bfseries 77} (1996) 2790}
  [\href{https://arxiv.org/abs/cond-mat/9604144}{{\ttfamily
  cond-mat/9604144}}].

\bibitem{PhysRevLett.83.3069}
A.W.~Sandvik, \emph{Multichain mean-field theory of quasi-one-dimensional
  quantum spin systems},
  \href{https://doi.org/10.1103/PhysRevLett.83.3069}{\emph{Phys. Rev. Lett.}
  {\bfseries 83} (1999) 3069}
  [\href{https://arxiv.org/abs/cond-mat/9904218}{{\ttfamily
  cond-mat/9904218}}].

\bibitem{Ramos_PRB_2020}
F.B.~Ramos, M.~Lencs\'es, J.C.~Xavier and R.G.~Pereira, \emph{Confinement and
  bound states of bound states in a transverse-field two-leg {Ising} ladder},
  \href{https://doi.org/10.1103/PhysRevB.102.014426}{\emph{Phys. Rev. B}
  {\bfseries 102} (2020) 014426}
  [\href{https://arxiv.org/abs/arXiv:2005.03145}{{\ttfamily
  arXiv:2005.03145}}].

\bibitem{Rutkevich_2010}
S.B.~Rutkevich, \emph{On the weak confinement of kinks in the one-dimensional
  quantum ferromagnet {CoNb}$_2${O}$_6$},
  \href{https://doi.org/10.1088/1742-5468/2010/07/P07015}{\emph{J. Stat. Mech.:
  Theory Exp} {\bfseries 2010} (2010) P07015}
  [\href{https://arxiv.org/abs/arXiv:1003.5654}{{\ttfamily arXiv:1003.5654}}].

\bibitem{jianda_E8_2014}
J.~Wu, M.~Kormos and Q.~Si, \emph{Finite-temperature spin dynamics in a
  perturbed quantum critical {I}sing chain with an ${E}_{8}$ symmetry},
  \href{https://doi.org/10.1103/PhysRevLett.113.247201}{\emph{Phys. Rev. Lett.}
  {\bfseries 113} (2014) 247201}
  [\href{https://arxiv.org/abs/arXiv:1403.7222}{{\ttfamily arXiv:1403.7222}}].

\bibitem{coldea_quantum_2010}
R.~Coldea et~al., \emph{Quantum {Criticality} in an {Ising} {Chain}:
  {Experimental} {evidence} for {emergent} {$E_8$} {symmetry}},
  \href{https://doi.org/10.1126/science.1180085}{\emph{Science} {\bfseries 327}
  (2010) 177} [\href{https://arxiv.org/abs/arXiv:1103.3694}{{\ttfamily
  arXiv:1103.3694}}].

\bibitem{E8}
H.~Zou et~al., \emph{{$E_8$} spectra of quasi-one-dimensional antiferromagnet
  {BaCo}$_2${V}$_2${O}$_8$ under transverse field},
  \href{https://doi.org/10.1103/PhysRevLett.127.077201}{\emph{Phys. Rev. Lett.}
  {\bfseries 127} (2021) 077201}
  [\href{https://arxiv.org/abs/arXiv:2005.13302}{{\ttfamily
  arXiv:2005.13302}}].

\bibitem{zhang_e8_observation_2020}
Z.~Zhang et~al., \emph{Observation of {$E_8$} particles in an {Ising} chain
  antiferromagnet},
  \href{https://doi.org/10.1103/PhysRevB.101.220411}{\emph{Phys. Rev. B}
  {\bfseries 101} (2020) 220411}
  [\href{https://arxiv.org/abs/arXiv:2005.13772}{{\ttfamily
  arXiv:2005.13772}}].

\bibitem{wang_tfic_quantum_2018}
Z.~Wang et~al., \emph{Quantum {criticality} of an {Ising}-like {spin}-$1/2$
  {antiferromagnetic} {chain} in a {transverse} magnetic field},
  \href{https://doi.org/10.1103/PhysRevLett.120.207205}{\emph{Phys. Rev. Lett.}
  {\bfseries 120} (2018) 207205}
  [\href{https://arxiv.org/abs/arXiv:1805.02392}{{\ttfamily
  arXiv:1805.02392}}].

\bibitem{PhysRevB.96.054423}
A.K.~Bera et~al., \emph{Spinon confinement in a quasi-one-dimensional
  anisotropic heisenberg magnet},
  \href{https://doi.org/10.1103/PhysRevB.96.054423}{\emph{Phys. Rev. B}
  {\bfseries 96} (2017) 054423}
  [\href{https://arxiv.org/abs/arXiv:1705.01259}{{\ttfamily
  arXiv:1705.01259}}].

\bibitem{WANG20242974}
X.~Wang et~al., \emph{Spin dynamics of the {$E_8$} particles},
  \href{https://doi.org/https://doi.org/10.1016/j.scib.2024.07.040}{\emph{Science
  Bulletin} {\bfseries 69} (2024) 2974}.

\bibitem{Kormos2017}
M.~Kormos, M.~Collura, G.~Tak{\'a}cs and P.~Calabrese, \emph{Real-time
  confinement following a quantum quench to a non-integrable model},
  \href{https://doi.org/10.1038/nphys3934}{\emph{Nat. Phys.} {\bfseries 13}
  (2017) 246} [\href{https://arxiv.org/abs/1604.03571}{{\ttfamily
  1604.03571}}].

\bibitem{PhysRevB.110.195101}
X.~Wang, M.~Oshikawa, M.~Kormos and J.~Wu, \emph{Magnetization oscillations in
  a periodically driven transverse field ising chain},
  \href{https://doi.org/10.1103/PhysRevB.110.195101}{\emph{Phys. Rev. B}
  {\bfseries 110} (2024) 195101}
  [\href{https://arxiv.org/abs/2408.13725}{{\ttfamily 2408.13725}}].

\bibitem{geometry}
X.~Wang, X.~He and J.~Wu, \emph{Many-body quantum geometry in time-dependent
  quantum systems with emergent quantum field theory instantaneously},
  \href{https://arxiv.org/abs/arXiv: 2503.18396}{{\ttfamily arXiv:
  2503.18396}}.

\bibitem{Pomponio:2024aiy}
O.~Pomponio, A.~Krasznai and G.~Tak\'acs, \emph{{Confinement and false vacuum
  decay on the Potts quantum spin chain}},
  \href{https://arxiv.org/abs/arXiv:2410.03382}{{\ttfamily arXiv:2410.03382}}.

\bibitem{SciRep2021}
J.~{Vovrosh} and J.~{Knolle}, \emph{{Confinement and entanglement dynamics on a
  digital quantum computer}},
  \href{https://doi.org/10.1038/s41598-021-90849-5}{\emph{Sci. Rep.} {\bfseries
  11} (2021) 11577} [\href{https://arxiv.org/abs/arXiv:2001.03044}{{\ttfamily
  arXiv:2001.03044}}].

\bibitem{Tan:2019kya}
W.L.~Tan et~al., \emph{{Domain-wall confinement and dynamics in a quantum
  simulator}}, \href{https://doi.org/10.1038/s41567-021-01194-3}{\emph{Nat.
  Phys.} {\bfseries 17} (2021) 742}
  [\href{https://arxiv.org/abs/arXiv:1912.11117}{{\ttfamily
  arXiv:1912.11117}}].

\bibitem{ZamE8}
A.B.~Zamolodchikov, \emph{{Integrals of motion and S matrix of the (scaled)
  $T=T_c$ {Ising} model with magnetic field}},
  \href{https://doi.org/10.1142/S0217751X8900176X}{\emph{Int. J. Mod. Phys. A}
  {\bfseries 4} (1989) 4235}.

\bibitem{zamolodchikov1987higher}
A.B.~Zamolodchikov, \emph{Higher order integrals of motion in two-dimensional
  models of the field theory with a broken conformal symmetry}, {\emph{JETP
  Lett.} {\bfseries 46} (1987) 160}.

\bibitem{Zamolodchikov:1989hfa}
A.B.~Zamolodchikov, \emph{{Integrable field theory from conformal field
  theory}}, {\emph{Adv. Stud. Pure Math.} {\bfseries 19} (1989) 641}.

\bibitem{sachdev_2011}
S.~Sachdev, \emph{Quantum Phase Transitions}, Cambridge University Press, 2~ed.
  (2011).

\bibitem{pfeuty_one-dimensional_1970}
P.~Pfeuty, \emph{The one-dimensional {Ising} model with a transverse field},
  \href{https://doi.org/10.1016/0003-4916(70)90270-8}{\emph{Ann. Phys.}
  {\bfseries 57} (1970) 79}.

\bibitem{coupleCFT}
A.~LeClair, A.~Ludwig and G.~Mussardo, \emph{Integrability of coupled conformal
  field theories},
  \href{https://doi.org/https://doi.org/10.1016/S0550-3213(97)00724-4}{\emph{Nucl.
  Phys. B} {\bfseries 512} (1998) 523}
  [\href{https://arxiv.org/abs/hep-th/9707159}{{\ttfamily hep-th/9707159}}].

\bibitem{QFTising}
J.B.~Zuber and C.~Itzykson, \emph{Quantum field theory and the two-dimensional
  {I}sing model}, \href{https://doi.org/10.1103/PhysRevD.15.2875}{\emph{Phys.
  Rev. D} {\bfseries 15} (1977) 2875}.

\bibitem{2isingboson}
D.~Boyanovsky, \emph{Field theory of the two-dimensional {I}sing model:
  Conformal invariance, order and disorder, and bosonization},
  \href{https://doi.org/10.1103/PhysRevB.39.6744}{\emph{Phys. Rev. B}
  {\bfseries 39} (1989) 6744}.

\bibitem{dark}
Y.~Gao, X.~Wang, N.~Xi, Y.~Jiang, R.~Yu and J.~Wu, \emph{Spin dynamics and dark
  particle in a weak-coupled quantum ising ladder with
  ${\mathcal{d}}_{8}^{(1)}$ spectrum},
  \href{https://doi.org/10.1103/PhysRevB.111.L201117}{\emph{Phys. Rev. B}
  {\bfseries 111} (2025) L201117}
  [\href{https://arxiv.org/abs/arXiv:2402.11229}{{\ttfamily
  arXiv:2402.11229}}].

\bibitem{Xning}
N.~Xi et~al., \emph{Emergent $\mathcal{D}_{8}^{(1)}$ mass spectrum in
  {CoNb}$_2${O}$_6$},  \href{https://arxiv.org/abs/arXiv:2403.10785}{{\ttfamily
  arXiv:2403.10785}}.

\bibitem{thermal}
Y.~Gao, J.~Yang, H.~Lin, R.~Yu and J.~Wu, \emph{Thermally activated detection
  of dark particles in a weakly coupled quantum ising ladder},
  \href{https://doi.org/10.1103/PhysRevB.111.L241105}{\emph{Phys. Rev. B}
  {\bfseries 111} (2025) L241105} [\href{https://arxiv.org/abs/arXiv:
  2406.15024}{{\ttfamily arXiv: 2406.15024}}].

\bibitem{rydberg}
B.W.~Shore, \emph{Coherent manipulation of atoms using laser light},
  {\emph{Acta Phys. Slovaca} {\bfseries 58} (2008) 243}.

\bibitem{DELFINO1996}
G.~Delfino, G.~Mussardo and P.~Simonetti, \emph{Non-integrable quantum field
  theories as perturbations of certain integrable models},
  \href{https://doi.org/https://doi.org/10.1016/0550-3213(96)00265-9}{\emph{Nucl.
  Phys. B} {\bfseries 473} (1996) 469}
  [\href{https://arxiv.org/abs/hep-th/9603011}{{\ttfamily hep-th/9603011}}].

\bibitem{DELFINO2006}
G.~Delfino, P.~Grinza and G.~Mussardo, \emph{Decay of particles above threshold
  in the {Ising} field theory with magnetic field},
  \href{https://doi.org/https://doi.org/10.1016/j.nuclphysb.2005.12.024}{\emph{Nucl.
  Phys. B} {\bfseries 737} (2006) 291}
  [\href{https://arxiv.org/abs/hep-th/0507133}{{\ttfamily hep-th/0507133}}].

\bibitem{Lieb1961}
E.~Lieb, T.~Schultz and D.~Mattis, \emph{Two soluble models of an
  antiferromagnetic chain},
  \href{https://doi.org/https://doi.org/10.1016/0003-4916(61)90115-4}{\emph{Ann.
  Phys.} {\bfseries 16} (1961) 407}.

\bibitem{Berg_1979}
B.~Berg, M.~Karowski and P.~Weisz, \emph{Construction of {Green's} functions
  from an exact {$S$} matrix},
  \href{https://doi.org/10.1103/PhysRevD.19.2477}{\emph{Phys. Rev. D}
  {\bfseries 19} (1979) 2477}.

\bibitem{BABELON1992113}
O.~Babelon and D.~Bernard, \emph{From form factors to correlation functions:
  The {Ising} model},
  \href{https://doi.org/https://doi.org/10.1016/0370-2693(92)91964-B}{\emph{Phys.
  Lett. B} {\bfseries 288} (1992) 113}
  [\href{https://arxiv.org/abs/hep-th/9206003}{{\ttfamily hep-th/9206003}}].

\bibitem{Bars:1977ud}
I.~Bars and M.B.~Green, \emph{Poincare and gauge invariant two-dimensional
  {QCD}}, \href{https://doi.org/10.1103/PhysRevD.17.537}{\emph{Phys. Rev. D}
  {\bfseries 17} (1978) 537}.

\bibitem{yurov1990}
V.P.~Yurov and A.B.~Zamolochikov, \emph{Truncated comformal space approach to
  scaling {Lee-Yang} model},
  \href{https://doi.org/10.1142/S0217751X9000218X}{\emph{Int. J. Mod. Phys. A}
  {\bfseries 05} (1990) 3221}.

\bibitem{yurov1991}
V.~Yurov and A.~Zamolochikov, \emph{Truncated-fermionic-space approach to the
  critical {2D} {Ising} model with magnetic field},
  \href{https://doi.org/10.1142/S0217751X91002161}{\emph{Int. J. Mod. Phys. A}
  {\bfseries 06} (1991) 4557}.

\bibitem{banks}
T.~Banks, D.~Horn and H.~Neuberger, \emph{Bosonization of the {SU(N)}
  {Thirring} models},
  \href{https://doi.org/https://doi.org/10.1016/0550-3213(76)90127-9}{\emph{Nucl.
  Phys. B} {\bfseries 108} (1976) 119}.

\bibitem{GINSPARG1988153}
P.~Ginsparg, \emph{Curiosities at c = 1},
  \href{https://doi.org/https://doi.org/10.1016/0550-3213(88)90249-0}{\emph{Nuclear
  Physics B} {\bfseries 295} (1988) 153}.

\bibitem{BASEILHAC2001607}
P.~Baseilhac, \emph{One-point functions in integrable coupled minimal models},
  \href{https://doi.org/https://doi.org/10.1016/S0550-3213(00)00632-5}{\emph{Nucl.
  Phys. B} {\bfseries 594} (2001) 607}
  [\href{https://arxiv.org/abs/hep-th/0005161}{{\ttfamily hep-th/0005161}}].

\bibitem{direct}
B.~Schroer and T.~Truong, \emph{Direct construction of the quantum field
  operators of the {D} = 2 {I}sing model},
  \href{https://doi.org/https://doi.org/10.1016/0370-2693(78)90821-3}{\emph{Phys.
  Lett. B} {\bfseries 73} (1978) 149}.

\bibitem{bookBosonization}
A.O.~Gogolin, A.A.~Nersesian and A.M.~Tsvelik, \emph{Bosonization and strongly
  correlated systems}, Cambridge University Press, Cambridge (2004).

\bibitem{Feverati:2006ni}
G.~Feverati, K.~Graham, P.A.~Pearce, G.Z.~Toth and G.~Watts, \emph{{A
  Renormalisation group for the truncated conformal space approach}},
  \href{https://doi.org/10.1088/1742-5468/2008/03/P03011}{\emph{J. Stat. Mech.}
  {\bfseries 0803} (2008) P03011}
  [\href{https://arxiv.org/abs/hep-th/0612203}{{\ttfamily hep-th/0612203}}].

\bibitem{Lencses:2015bpa}
M.~Lencses and G.~Takacs, \emph{{Confinement in the q-state Potts model: an
  RG-TCSA study}}, \href{https://doi.org/10.1007/JHEP09(2015)146}{\emph{JHEP}
  {\bfseries 09} (2015) 146}
  [\href{https://arxiv.org/abs/1506.06477}{{\ttfamily 1506.06477}}].

\bibitem{KAROWSKI1978455}
M.~Karowski and P.~Weisz, \emph{Exact form factors in (1 + 1)-dimensional field
  theoretic models with soliton behaviour},
  \href{https://doi.org/https://doi.org/10.1016/0550-3213(78)90362-0}{\emph{Nucl.
  Phys. B} {\bfseries 139} (1978) 455}.

\bibitem{Smirnov}
F.A.~Smirnov, \emph{Form Factors in Completely Integrable Models of Quantum
  Field Theory}, WORLD SCIENTIFIC (1992).

\bibitem{Lukyanov1997}
S.~Lukyanov, \emph{Form factors of exponential fields in the sine-{Gordon}
  model},
  \href{https://doi.org/https://doi.org/10.1142/S0217732397002673}{\emph{Mod.
  Phys. Lett. A} {\bfseries 12} (1997) 2543}
  [\href{https://arxiv.org/abs/hep-th/9703190}{{\ttfamily hep-th/9703190}}].

\bibitem{Mussardobook}
G.~Mussardo, \emph{{Statistical field theory: an introduction to exactly solved
  models in statistical physics; 1st ed.}}, Oxford graduate texts, Oxford Univ.
  Press, New York, NY (2010).

\bibitem{BAJNOK2001503}
Z.~Bajnok, L.~Palla, G.~Tak\'acs and F.~W\'agner, \emph{Nonperturbative study
  of the two-frequency sine-{Gordon} model},
  \href{https://doi.org/https://doi.org/10.1016/S0550-3213(01)00067-0}{\emph{Nucl.
  Phys. B} {\bfseries 601} (2001) 503}
  [\href{https://arxiv.org/abs/hep-th/0008066}{{\ttfamily hep-th/0008066}}].

\bibitem{POZSGAY2008}
B.~Pozsgay and G.~Tak\'acs, \emph{Form factors in finite volume {I}: Form
  factor bootstrap and truncated conformal space},
  \href{https://doi.org/https://doi.org/10.1016/j.nuclphysb.2007.06.027}{\emph{Nucl.
  Phys. B} {\bfseries 788} (2008) 167}
  [\href{https://arxiv.org/abs/arXiv:0706.1445}{{\ttfamily arXiv:0706.1445}}].

\bibitem{marton2019}
K.~H\'ods\'agi, M.~Kormos and G.~Tak\'acs, \emph{Perturbative post-quench
  overlaps in quantum field theory},
  \href{https://doi.org/10.1007/JHEP08(2019)047}{\emph{J. High Energy Phys.}
  {\bfseries 2019} (2008) 47}
  [\href{https://arxiv.org/abs/arXiv:1905.05623}{{\ttfamily
  arXiv:1905.05623}}].

\end{thebibliography}\endgroup
\end{document}